\documentclass[acmsmall]{acmart}
\usepackage{graphics}
\usepackage{graphicx}
\usepackage{amsmath}
\usepackage{amsfonts}
\usepackage{cases}
\usepackage{setspace}
\usepackage{xspace}
\usepackage{multirow}
\usepackage{subfigure}
\usepackage{color}
\usepackage{url}
\usepackage{booktabs}
\usepackage{ftnxtra}
\usepackage{pbox}
\usepackage[linesnumbered,boxed,ruled,commentsnumbered]{algorithm2e}
\usepackage{stmaryrd}
\usepackage{mathrsfs}
\usepackage[american]{babel}
\usepackage{microtype}
\usepackage{listings}
\usepackage{xcolor}

\usepackage{color}
\usepackage{amsmath}
\usepackage{multirow}
\usepackage{colortbl}

\usepackage{adjustbox}

\usepackage{cases}
\usepackage{textcomp}

\definecolor{dkgreen}{rgb}{0,0.6,0}
\definecolor{gray}{rgb}{0.5,0.5,0.5}
\definecolor{mauve}{rgb}{0.58,0,0.82}
\usepackage{todonotes}

\newcommand{\toolname}{{\sf ACCENT}}

\usepackage{amsmath}
\usepackage{multirow}
\usepackage{cases}
\usepackage{textcomp}
\usepackage{wrapfig,lipsum}

\newcommand{\secondrev}[1]{\textcolor{black}{#1}}
 
\newcommand{\revision}[1]{\textcolor{black}{#1}}
 
\AtBeginDocument{%
  \providecommand\BibTeX{{%
    \normalfont B\kern-0.5em{\scshape i\kern-0.25em b}\kern-0.8em\TeX}}}

\setcopyright{acmcopyright}
\copyrightyear{}
\acmYear{}
\acmDOI{}

\acmJournal{TOSEM}
\acmVolume{0}
\acmNumber{0}
\acmArticle{0}
\acmMonth{0}

\begin{document}

\title{Adversarial Robustness of Deep Code Comment Generation}

\author{Yu Zhou}
\email{zhouyu@nuaa.edu.cn}
\author{Xiaoqing Zhang}
\email{zhangxq@nuaa.edu.cn}
\author{Juanjuan Shen}
\email{shenjuanjuan@nuaa.edu.cn}
\affiliation{%
  \institution{Nanjing University of Aeronautics and Astronautics}
 \city{Nanjing}
  \country{China}
}

\author{Tingting Han}
\email{t.han@bbk.ac.uk}
\author{Taolue Chen}
\email{t.chen@bbk.ac.uk}
\authornote{Corresponding author.}
\affiliation{%
 \institution{Birkbeck, University of London}
 \city{London}
 \country{UK}}

\author{Harald Gall}
\email{gall@ifi.uzh.ch}
\affiliation{%
	\institution{University of Zurich}
	\city{Zurich}
	\country{Switzerland}
}

\renewcommand{\shortauthors}{Y. Zhou, et al.}

\begin{abstract}
Deep neural networks (DNNs) have shown remarkable performance in a variety of domains such as computer vision, speech recognition, and natural language processing. Recently they also have been applied to various software engineering tasks, typically involving processing source code. DNNs are well-known to be vulnerable to adversarial examples, i.e., fabricated inputs that could lead to various misbehaviors of the DNN model while being perceived as benign by humans. In this paper, we focus on the code comment generation task in software engineering and study the robustness issue of the DNNs when they are applied to this task. We propose {\toolname} (\textbf{A}dversarial \textbf{C}ode \textbf{C}omment g\textbf{EN}era\textbf{T}or), an identifier substitution approach to craft adversarial code snippets, which are syntactically correct and \secondrev{semantically close to} the original code snippet, but may mislead the DNNs to produce completely irrelevant code comments. In order to improve the robustness, {\toolname} also incorporates a novel training method, which can be applied to existing code comment generation models. We conduct comprehensive experiments to evaluate our approach by attacking the mainstream encoder-decoder architectures on two large-scale publicly available  datasets. The results show that {\toolname} efficiently produces stable attacks with functionality-preserving adversarial examples, and the generated examples have better transferability compared with the baselines. We also confirm, via experiments, the effectiveness in improving model robustness with our training method. 
\end{abstract}



\maketitle

\section{Introduction} \label{sect:intro}
Code comment generation aims to generate readable natural language descriptions of source code snippets, which plays an important role in facilitating program comprehension. 
Encouraged by the great success of deep learning methods in typical application areas 
such as computer vision and natural language processing, researchers have proposed deep neural network (DNN) based approaches for the code comment generation task \cite{Hu2018Summarizing,Yu2019Augmenting,2020A}, aiming to improve the quality of the generated comments. 

It is well-recognized that DNNs are not robust. In particular, adversarial examples, which can be crafted by adding small perturbations to benign inputs of the model, 
may easily fool DNNs \cite{Goodfellow2015Explaining,Papernot2016Practical}, or at least elicit large changes in the model output. This would greatly impede the usability of DNN models~\cite{ramakrishnan20}, since ideally 
the model should generate indistinguishable comments for similar code snippets. In other words, minor semantic-preserving perturbation of the code snippets should have a minimum side-effect on the generated comments. As a result, when  neural networks are adopted, improving their robustness has become indispensable. 

Adversarial examples have been shown  to be an effective way of assessing and improving the robustness of neural networks. In light of this, recently the deep learning community has seen a wide variety of methods to generate adversarial examples, 
especially for image classification \cite{Goodfellow2015Explaining,Carlini2016Towards} and some NLP tasks \cite{Papernot2016Crafting,2018HotFlip,Zou2020}. Likewise, when applying DNNs to programming and software engineering tasks, it is also vital to improve the robustness of the model, which demands effective and efficient ways to generate adversarial examples. However,  this is considerably more challenging for the source code of programming languages. One of the reasons is that it must satisfy various syntactic and semantic constraints, which are more stringent than the image or NLP cases. For instance, the syntactic constraint stipulates that the adversarial code snippet must be compilable and executable, whereas the semantic constraint stipulates that it must preserve the ``meaning'' of the original code. Nonetheless, the perturbations reveal the weakness of the model only if they do not change the input so significantly but can legitimately result in changes in the expected output. Another source of difficulty lies in that, comparatively speaking, generating adversarial images is usually much easier because, fundamentally it is a continuous optimization problem where powerful, gradient-based techniques can be utilized.  
In contrast, the adversarial program is of discrete nature.   
Notice that, when applying deep learning methods to source code, the program snippet is usually embedded into a vector space, giving rise to a continuous representation. However, in general, there is no correspondence between the perturbed representation and the valid tokens in the code snippet, which rules out a straightforward adaptation of the current approaches in image classification to the domain of programs. 

The current task is more akin to the NLP domain as both are dealing with discrete texts. The key difference is that in the case of programs, one has to consider the rigid grammar imposed on programs; fundamentally a programming language is an abstract 
language. In other words, the adversarial perturbed code must be compilable and semantically equivalent for which natural languages (such as English) are much more liberal and easier to achieve. In contrast, it becomes harder to synthesize adversarial examples for source code when applying NLP methods directly. 

In this paper, we propose a novel approach {\toolname} (\textbf{A}dversarial \textbf{C}ode \textbf{C}omment g\textbf{EN}era\textbf{T}or) to generate adversarial examples and improve the robustness of neural networks for the code comment generation task. In a nutshell, we identify the importance of different identifiers appearing in the code snippet and rename them iteratively without breaking the syntactic structure and semantic of the code snippet. Furthermore, we adopt a new training method to improve the robustness of code comment generation models.

\begin{figure}[!t]
	\centering
	\includegraphics[scale=0.6]{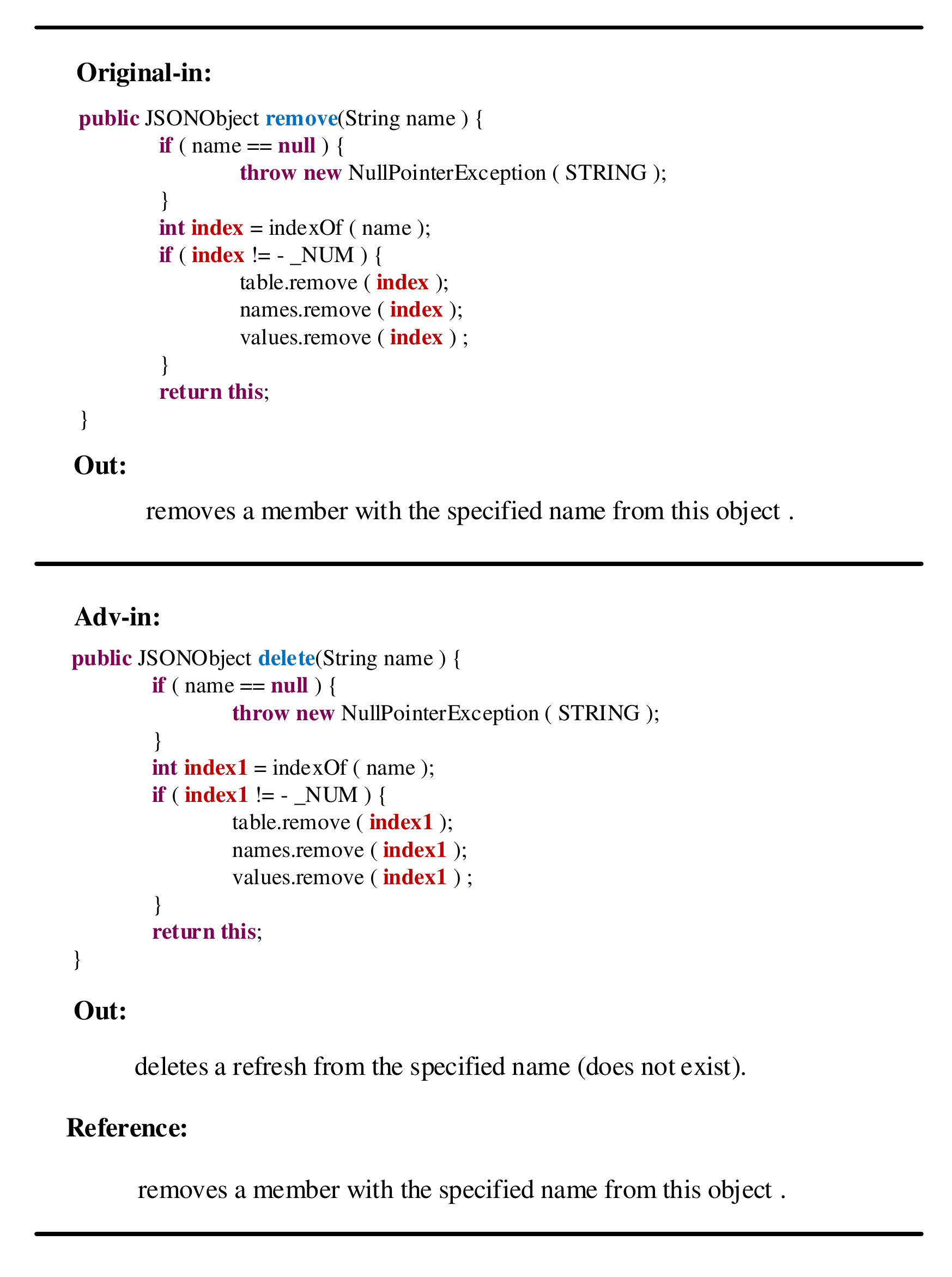}
	\caption{An adversarial attack example for a comment generation model}
	\label{fig:sample}
\end{figure}

Figure~\ref{fig:sample} exemplifies an adversarial example, where `in' and `out' refer to the input code snippet and the generated comment by a comment generator based on the Transformer architecture~\cite{2020A}. This example substitutes the function name `remove' with `delete' and `index' with `index1', which are syntactically correct and clearly does not change the semantic or the functionality of the code (cf.\ adv-in), so should have very similar comments. However, rather surprisingly, for this seemingly innocent new code snippet, the comment ``deletes a refresh from the specified name (does not exist).\secondrev{"} is generated, which is completely irrelevant and indeed very distant from the reference comment.


		
	

We carry out evaluations to assess the effectiveness of our approach. For the dataset, we use the publicly available Java source code dataset \cite{Hu2018Summarizing} which was extracted from GitHub,\footnote{https://github.com/} and Python dataset which was extracted from \cite{2018Improving}. We consider \revision{five} sequence-to-sequence (seq2seq) models for comment generation for which representative work of various architectures is selected. 
The experimental results show that {\toolname} is capable of attacking all different models and is beneficial to improve the adversarial robustness without jeopardizing the performance (i.e., the quality of the generated comments). 

Our main contributions are summarized as follows.
\begin{itemize}
	\item We propose a novel approach to assess and improve the robustness of neural source code models for the comment generation task, including both adversarial examples generation and novel training methods. To the best of our knowledge, it represents one of the first work addressing the model robustness of such a task. 
	\item We conduct comprehensive experiments to demonstrate the effectiveness of our approach, which also confirms the transferability of the generated adversarial examples, crucial for black-box attacks.
	\item We make the implementation of our approach, as well as the datasets publicly available,\footnote{https://github.com/zhangxq-1/ACCENT-repository}  which not only can facilitate the replication of our work, but also provides potential usage for related software engineering research and practice.
\end{itemize}

\smallskip

\noindent\emph{Structure of the paper.} \revision{The remainder of the paper is organized as follows. Section~\ref{sect:prel} introduces the
background. Section~\ref{sect:appr} describes the technical details of our approach, and Section~\ref{sect:eval} presents the
experimental results. Discussions are given in Section~\ref{Threats to Validity}, followed by a discussion of the related
work in Section~\ref{sect:rel}. We conclude the paper and outline future research plans in Section~\ref{sect:conc}.
}

\section{Background} \label{sect:prel}

\subsection{Source Code Comment Generation}
Code comment generation is a typical software engineering task. Here both the code and the generated comment are regarded as sequences of tokens that can be represented by vectors, for which sequence-to-sequence (seq2seq) models are suitable and commonly adopted. In a nutshell, the seq2seq model turns one sequence into another one utilizing a recurrent neural network (RNN) or variants thereof, such as long short-term memory (LSTM) or gated recurrent unit (GRU) models, to avoid the problem of vanishing gradient. 
Typically, the model is based on the encoder-decoder architecture where both encoder and decoder are neural networks; the former turns each item into a corresponding hidden vector containing the item and its context, and the latter reverses the process, turning the vector into an output item, using the previous output as the input context. In addition to the classic RNN-based approaches \cite{Bahdanau2014Neural}, recent developments include various attention mechanisms \cite{2017Attention} which allow the decoder to look at the input sequence selectively rather than generate a single vector which stores the entire context. A typical of example of the models with attentions is Transformer and BERT.

\subsection{Adversarial Attacks}
Adversarial attacks can be described as the process that, given the original input $x$, finds an adversarial perturbation $\delta$ such that $x+\delta$ can dramatically degrade the model's performance. 
Adversarial attacks can be conducted in both white-box and black-box manners depending on the attacker's knowledge on the model. For white-box attacks \cite{Papernot2016Crafting,2017Towards}, attackers have full access to the target model, e.g., the architecture and parameters. For black-box attacks \cite{2018Black}, they have no or little knowledge about the target model. From another perspective, according to the purpose of the attacker, there are targeted or non-targeted adversarial attacks. Take the classification model as an example, attackers purposefully mislead the model to a selected label in the targeted attack, while  they only aim at fooling the model in the non-targeted attack.

%
One can adapt the adversarial attacks to the code comment generation setting in a rather straightforward way. Given a (well-trained) code comment generation model $M$, 
an adversarial example can be  generated by identifying a perturbation $\delta$ that maximizes the model degradation $L_{adv}$. 
Formally, $x^*=x+\delta$, where 
\[ 
\delta:=\underset{||\delta||_p\leq \epsilon}{\arg\max}\left\{L_{adv}(x+\delta)-\lambda C(\delta)\right\} 
\]
Here, $C(\delta)$ captures the semantic and syntactic constraints; $\lambda$ is the regularization penalty; $||\delta||_p$ represents the constraint on the perturbation $\delta$. Note this seemingly simple formulation does not lend itself to efficient solutions; it merely provides a conceptual framework. 
%
%

\smallskip
\noindent\textbf{Transferability of adversarial examples.} The transferability of adversarial examples has been widely exploited in adversarial attacks, which refers to the phenomenon that examples generated on one model can also be used to attack other models for similar tasks \cite{Goodfellow2015Explaining,2018Distorting}. Transferability is an important property reflecting the generalizability of the attack method, i.e., the higher the transferability of adversarial examples, the better the generalization ability of the attack method.

\subsection{Defense and Robustness} \label{sect:dr}
To thwart adversarial attacks, various defense methods have been proposed to protect DNN models. In general, defense methods can be classified into two categories: detection and model enhancement \cite{2019Towards}. For \secondrev{the} former, defenders try to detect adversarial examples so can shield  the model from them. For the latter, the main task is to train the model to enhance its  robustness. Among others, adversarial training \cite{Goodfellow2015Explaining} is a widely adopted model enhancement approach, which has been successfully applied to image processing \cite{2015Learning} and NLP \cite{2018TextBugger,2018Black,2018Generating} domains. In a nutshell, it mixes adversarial examples with the original dataset to synthesize a new dataset, which is used to re-train the model.


\secondrev{
Note that in literature there are a number of variants of adversarial training, which is usually used as an umbrella term to refer to a family of training methods that  utilize adversarial examples to improve the robustness of deep learning models. For instance, Madry et al. \cite{MadryMSTV18} formulate the adversarial training as a min-max optimization problem, which is challenging to solve. In this paper, we 
instead pursue a lightwight adversarial training method in Section~\ref{sect32}.
}


The robustness of the model has a far-reaching influence on deep learning in, for example, representation learning and model interpretability. Ilyas et al. \cite{2019Adversarial} claim that adversarial vulnerability is caused by non-robust features. DNN models are vulnerable to attacks because of the well generalizing non-robust features in the data. Although robust features and non-robust features are both useful, a robust model should learn  the robust features, rather than non-robust ones. Our work contributes to the understanding of the model robustness for a new application domain, i.e., software engineering.  


\section{Our Approach} \label{sect:appr}
\begin{figure}[!t]
	\centering
	\includegraphics[scale=0.43]{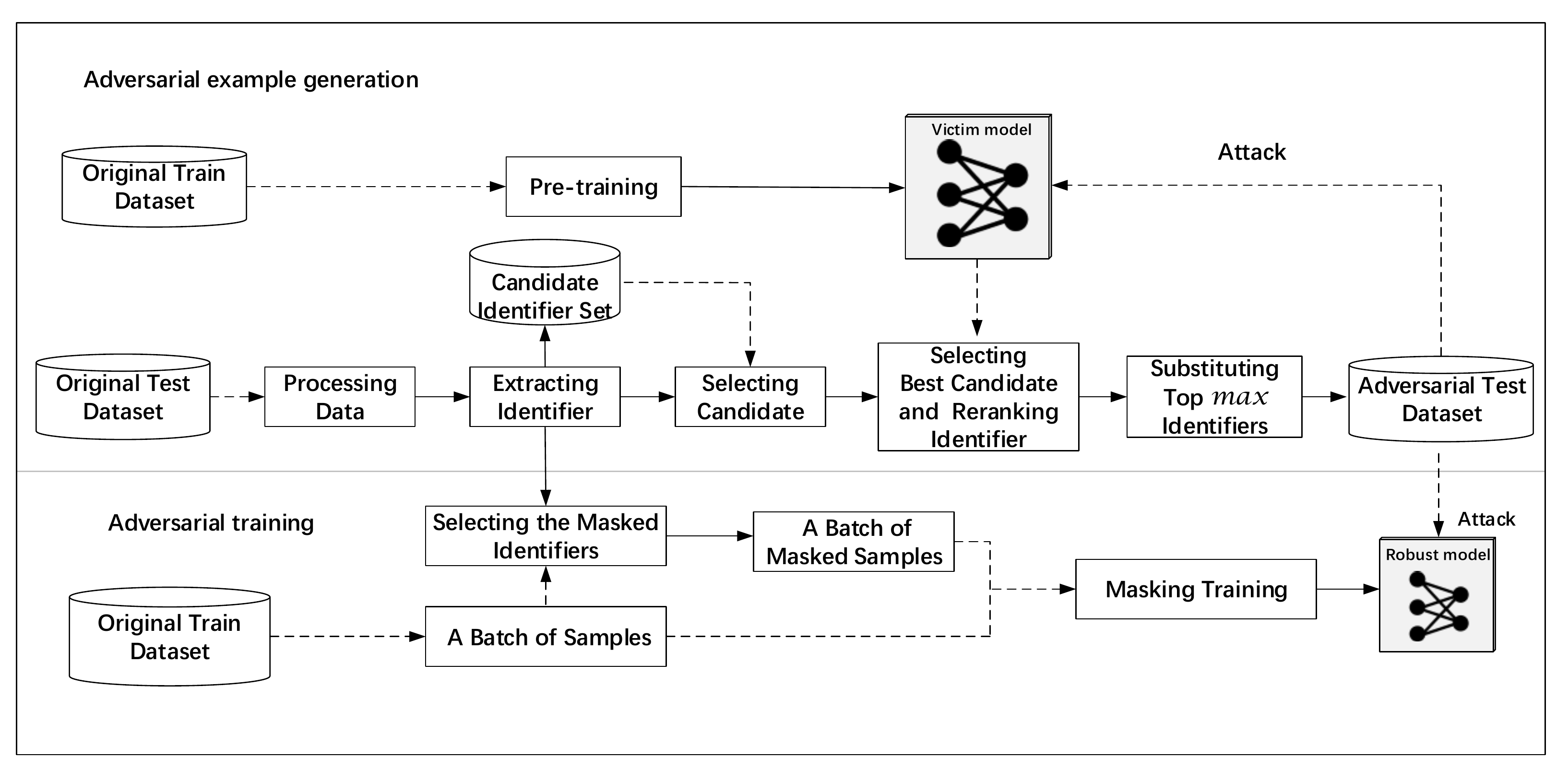}
	\caption{The workflow of \toolname}
	\label{fig:workflow}
\end{figure}
%
%

The overview of {\toolname} is given in Figure~\ref{fig:workflow}. There are mainly two parts in {\toolname}, i.e., adversarial examples generation and adversarial training. For \revision{the} former, source code in the original test data-set goes through a series of processing steps to generate the best candidate identifiers to substitute  as adversarial examples.  For the latter, the original training data and the masked data are used together for the adversarial training.  The details of these two parts are described in Section~\ref{sect31} and Section~\ref{sect32} respectively.

\begin{algorithm}[htb] 
	\caption{Adversarial Example Generation Algorithm} \label{alg:Framwork} 
	\KwIn{
		Code Comment Generation Model $M$;\\
	Code Comment Generation	DataSet $D$, where $(p,com)\in D$, 
	 $p$ is the original program snippet and $com$ is the comment;\\
		Max Substitute Number $\max$;\\
		Candidate Identifier Number $K$;\\}
	\KwOut{
		Adversarial DataSet $D_{adv}$;}
	Initialize: Candidate Identifier Set $V\gets \emptyset$, Adversarial DataSet $D_{adv}\gets \emptyset$\;
	\For{each $(p,com)\in D$}
	{
		$V \gets V \cup$ \{$w \mid$ $w$ is an identifier and $w$ is defined in $p$\};
	}
	Training Identifier Embedding $Embed$\;
	\For {each $(p,com)\in D$}
	{Extract the identifier set $V_{p}$ for $p$ by $V_{p}\gets \{ w$ $\mid$ $w$ is an identifier and $w$ is defined in $p$\}\;
		\For {each $w \in V_{p}$}
		{Select $K$ candidate substitute identifiers $L_w$$\subseteq V-V_{p}$ for the identifier $w$ based on the cosine similarity\;
		$w^* \gets \underset{w'\in L_w}{\arg\max} \left\{score(p)-score(p[w\leftarrow w'])\right\}$\;
			Extract the embedding of identifier $w$ from $Embed$ and the embedding of $p$ from encoder\;
			Calculate the identifier saliency $S(p,w)$\;
			Calculate $H(p,p^*,w)$\;}
		For $w \in V_{p}$, reorder $w$ according to $H(p,p^*,w)$ in descending order\;
		\For {$index\gets 1$ to $\max$}
		{Generate $p_{adv}$ by replacing $w$ with $w^*$;}
		$D_{adv} \gets$ $D_{adv} \cup \{(p_{adv},com)\}$}
	\Return $D_{adv}$\; 
\end{algorithm}

\subsection{Adversarial Attack} \label{sect31}
For the adversarial attack, we mainly consider two types of programmer-defined identifiers, i.e., single-letter and non-single-letter identifiers. For the former, we simply change it to a different letter randomly. For the latter, we adopt a black-box, non-target search-based method to generate adversarial examples. We first extract identifiers from \secondrev{all the program in the dataset to build up a candidate identifier set. For each identifier in the program, we select the nearest $K$ identifiers from the candidate set according to the cosine similarity to form a sub-candidate set,} from which the best candidate is identified based on its effect on the generated code comment. 
We then rank these candidate identifiers based on their contextual relation to the program. Finally, we generate adversarial examples by replacing the identifier with its best candidate according to the order determined in the ranking. In the sequel, we elaborate these steps. 

\paragraph{Step 1: Identifier Extraction (``Extracting Identifier'' in Figure~\ref{fig:workflow})} The first step is to extract identifiers from program snippets and build a candidate identifier set (cf.\ Line 2-3 in Algorithm~\ref{alg:Framwork}). 
\secondrev{Since the functionality of a program snippet does not depend on the programmer-defined  identifiers, changing them should preserve the execution of the program, which is more likely to preserve the semantics of the program. As a result, we choose these identifiers such as method names and variable names as our target identifiers to be substituted.}

To facilitate the extraction, we exploit abstract syntax trees (ASTs). 
We use Javalang\footnote{https://github.com/c2nes/javalang \label{javalang}} to obtain ASTs for Java code, and the ast\footnote{https://docs.python.org/3/library/ast.html \label{ast}} lib  for Python code. The identifiers are then extracted based on the node types in the ASTs; afterwards, they are put into an identifier candidate set $V$.

\paragraph{Step 2: Candidate Selection (``Selecting Candidate'' in Figure~\ref{fig:workflow}).} The size of the extracted identifier set is usually extremely large. To speed up the search for the optimal substitution identifier, for each identifier $w$ in the program $p$, we construct a subset $L_{w}\subseteq V$, which contains
$K$ identifiers that have the shortest distance to  $w$ (cf.\ Line 9 in Algorithm~\ref{alg:Framwork}). Note that here $K$ is a hyper-parameter. (In our experiment we set $K=5$.) 
Each identifier in $L_{w}$ is then considered to be a candidate for the substitution of $w$. 
To obtain $L_{w}$, we train embeddings using word2vec \cite{mikolov2013distributed} with the skip-gram algorithm (cf.\ Line 5 in Algorithm~\ref{alg:Framwork}). The skip-gram algorithm is to construct word representations (i.e., word embedding) that are useful for predicting the surrounding words in a given corpus. Given a sequence of training words $w_1,\cdots, w_n$, the objective of the skip-gram is to maximize the average logarithm of the probability:
\[
\frac 1n \sum\limits_{t=1}^n\sum\limits_{-c\leq j \leq c,j\neq 0}\log p(w_{t+j}|w_t),  
\]
where $c$ is the training context. Note that we use the tokens split from the program snippet rather than identifiers solely as the training corpus, and then extract the embeddings of the identifier set obtained in the previous step. 
For each identifier $w$, 
we select the $K$ nearest identifiers according to the cosine distance, viz.,  
\[
L_{w}=top_K(cos(w,V')) 
\]
Here, $V'$ is the set of identifiers obtained by deleting the identifiers and formal parameters that appeared  
in  $V$, so we can make sure that the  program after substitution is compilable. 
Each identifier  in $L_{w}$ is then considered to be a candidate for the substitution of $w$.

Importantly, we adopt the cosine similarity in selecting the candidate replacement. The reason is, 
when the identifier in the original program is substituted by one in the candidate set $L_w$ to generate the adversarial examples, the program semantics should not be changed significantly (which implies that the generated comments should be similar for a robust model). 

The following example shows that a naive approach would not serve the purpose. In this example, the original program is
 
{\color{black}\verb|float avg_velocity(float distance, float time) {return distance/time;}|}

\noindent\secondrev{When we replace identifiers, possibly the method name ``avg\_velocity" is replaced by ``density", and the arguments ``distance" and ``time" are replaced by ``mass" and ``volume" respectively. Namely, we obtain} 

{\color{black}\verb|float density(float mass, float volume) {return mass/volume;}|}

\noindent As one can argue easily, the resulting program is quite different from the original program in semantics and thus should have a different comment. In other words, it should \emph{not} be considered as an adversarial example. %
To rule out these cases, we adopt a \emph{constrained substitution} approach. Namely, we utilize the word embedding method (word2vec in our implementation) and cosine similarity to only allow those identifies which are semantically related to the original identifies to be replaced. In this way, the obtained code snippet would be close to the original one in semantics and would be functionality preserving, and, if its comment deviates from the original comment significantly, it should be regarded as a valid adversarial example. 



\paragraph{Step 3: Best Candidate Selection and Identifier Reranking (``Selecting Best Candidate and Reranking Identifier'' in Figure~\ref{fig:workflow})} For each identifier $w$ extracted from the program, we have obtained 
a candidate set $L_w$ that contains $K$ identifiers. Then we replace $w$ with each $w'$ in $L_w$ and calculate  the score change of the generated comment  after substitution (cf.\ Line 9 in Algorithm~\ref{alg:Framwork}). 
We define
\[
w^*= \underset{w'\in L_w}{\arg\max} \{score(p)-score(p[w\leftarrow w'])\} 
\]
where $p[w\leftarrow w']$ is the new program obtained by replacing $w$ with the candidate identifier $w'\in L_w$. 
%
In other words, $w^*$ is the one which causes the most significant change and is replaced by $w^*$ to generate a new program $p^*$. $score(p)$ is the output of the original deep code comment generation model by feeding the input $p$.
%
For the code comment generation task, we use the BLEU score as the metric for the generated comment in natural language. 

The change on the result between $p$ and $p^*$ represents the best attack effect that can be achieved after replacing $w$, i.e., $\Delta score_w^*=score(p)-score(p^*)$. 
For each identifier $w$, we iterate all candidate identifiers $w^*$ and calculate the corresponding $\Delta score_w^*$.

A program snippet usually contains multiple identifiers, and each identifier may have different levels of contextual relation to the original program. 
We then adopt identifier saliency 
to quantify the degree of the contextual relation between the identifiers and the original program, which will be used  to determine the identifier substitution order.  
The saliency of an identifier $w$  with respect to a program $p$, i.e., $S(p,w)$, is computed as $\cos(vec(w),vec(p))$ where 
\[
cos(vec(w),vec(p))=\frac{vec(w)\cdot vec(p)}{||vec(w)||\cdot ||vec(p)||},
\]
$vec(w)$ is the embedding of $w$, and  $vec(p)$ is the contextual encoder of the program $p$. Here, we train an independent encoder-decoder model based on a single-layer LSTM using the two publicly available datasets, and extract the output of the encoder as the embedding of $p$ (cf.\ Line 10 in Algorithm~\ref{alg:Framwork}).


For all $w$ extracted from $p$, we calculate the identifier saliency $S(p,w)$ to obtain a saliency vector \textbf{$S(p)$} (cf.\ Line 11 in Algorithm~\ref{alg:Framwork}).  Then, for each identifier, we consider the change after substitution $\Delta score_w^*$ and  the identifier saliency $S(p,w)$ 
to determine the  order of substitution. We define a score function $H(p,p^*,w)$ to score each identifier and sort all the identifiers in $p$ in descending order based on $H(p,p^*,w)$ (cf. Line 12 in Algorithm~\ref{alg:Framwork}). The score function $H(p,p^*,w_i)$ is defined as 
\[ 
H(p,p^*,w)=\left\{
\begin{aligned}
S(p,w)  \cdot \Delta score_w^* \quad  S(p,w)\neq 0,\Delta score_w^{*}\neq 0 \\
S(p,w) \cdot \beta ~~~~ \quad S(p,w)\neq 0, \Delta score_w^{*}= 0 \\
\Delta score_w^* \cdot \alpha~~~~~~ \quad S(p,w)=  0,\Delta score_w^{*}\neq 0 \\
0~~~~~~~~~~~~~~~~~~~~~~~~~~~~ \quad {o.w.}~~~~~~~~~~~~ \\
\end{aligned}
\right
.
\]
where $\alpha$ and $\beta$ $\in[0, 1]$ are the constant parameters.
%

The definition of the score function $H$ considers both the change of the model output after identifier substitution and the importance of the substituted identifier to the original program snippet. 
In particular, $S(p, w)$ focuses on describing the impact of the identifier $w$ on the original program snippet, while $\Delta score_w^*$ focuses on the impact on the model. In order to reduce the interference of the two metrics (i.e., to avoid the weighted score function vanishes when one of them vanishes), we simply take one of them when the other vanishes. 


\paragraph{Step 4: Adversarial Example Generation (``Substituting Top $max$ Identifiers'' in Figure~\ref{fig:workflow}).}
\secondrev{We reorder all identifiers according to $H(p,p^*,w)$ 
	and select the top $max$ identifiers to replace (cf.\ Line 15-18 in Algorithm~\ref{alg:Framwork}). }
To ensure that the program is compilable, we replace all occurrences where the identifier has appeared in the program. For example, if we replace `A' with `B' in 
``\verb|void f() {int A=1; A++;}|'', 
the new program becomes
``\verb|void f() {int B=1; B=B++;}|''.

\subsection{Robustness Improvement} \label{sect32}

Adversarial training 
aims to improve the robustness of deep learning models intrinsically. In the last few years, a variety of adversarial training methods have been proposed. In the sequel, we propose masked training, which considered to be a lightweight adversarial training method tailored to the code comment generation setting.
%


\begin{algorithm}[htb] 
	\caption{Masked Training Algorithm}\label{training} 
	
	\KwIn{
		Code Comment Generation DataSet $D$\;
		The number of Masked Identifiers $Count_{masked}$\;
		Hyperparameter 	 $\lambda$\;
		
	}
	\KwOut{
		Trained model $M_{masked}$\;}
	Initializing the model parameters $M_{masked}$ according to the original deep code comment generation model training method \;
	\For{batch $d$ of data  $\in D$}
	{   \For{$(p,com)\in d$}{
			Randomly mask $Count_{masked}$ identifiers;
		}
		Calculate origin loss: $L_{origin}(p, com)$\;
		Calculate masked loss: $L_{masked}(p', com)$\;
		Train the model $M_{masked}$ according to $\theta^\star \gets \arg\min_\theta {\left(\lambda*L_{origin}(p, com)+(1-\lambda)*L_{masked}((p', com)\right)}$
	}
	\Return $M_{masked}$\; 
\end{algorithm}

As mentioned in Section~\ref{sect:dr}, the low degree of robustness may be caused by the reliance on the so called non-robust features. As a result, the general idea of masked training is to reduce the dependence of the model on the non-robust features since any perturbations upon these features may cause great change on the output. Algorithm~\ref{training} illustrates the workflow of the method. Given source code $p$, we generate the corresponding masked code $p'$ (cf. Line 3-5 in Algorithm~\ref{training}), which is constructed by randomly replacing $k$ identifiers in $p$ by $<unk>$. The general  objective function for the masked training is defined as 
\[
\theta^\star= \arg\min_\theta L(p, com), 
\]
where $L(p, com)$ is the negative log-likelihood 
\[ 
L(p, com)=-\frac{1}{m} \sum\limits_{t=1}^m logP(com_t|com_{<t},p)
\]
In particular, we  employ two objective functions to improve the robustness of the model (cf. Line 6-8 in Algorithm~\ref{training}). Namely, 
$L_{origin}(p, com)$ which can guarantee good performance while keeping the stability of the model and 
$L_{masked}(p',com)$, which can guide the model to generate the output $com$ according to the masked input $p'$, making the output of the model independent of the identifiers. 

Formally, given a model and the training corpus, the masked training objective is
\[ 
\theta^\star=\arg\min_\theta \left(\lambda \cdot L_{origin}(p, com)+(1-\lambda)\cdot L_{masked}(p', com)\right), 
\]%
where $\lambda$ is a hyperparameter.


\section{Evaluation} \label{sect:eval}
\subsection{Experiment setup}

We conduct comprehensive experiments to demonstrate the effectiveness of  the proposed approach on the Java source code dataset \cite{Hu2018Summarizing} and the Python source code dataset  \cite{2018Improving}, which are widely adopted benchmarks for the code comment generation task.  The statistics of the two datasets are shown in Table~\ref{Table:1}. For the Java dataset, we follow the original work \cite{Hu2018Summarizing} which divided the 
\begin{wraptable}{r}{5cm}
	\begin{center}
		\caption{Statistics of datasets}
		\begin{tabular}{|c|c|c|}
			\hline
			Dataset    & Java  & Python \\ \hline
			Train     & 69,708 & 50,400  \\ \hline
			Validation & 8,714  & 13,248  \\ \hline
			Test       & 8,714  & 13,216  \\ \hline
		\end{tabular}
		\label{Table:1} 
	\end{center}
\end{wraptable}
examples into train dataset, validation dataset and test dataset in the ratio of 8:1:1. For the Python dataset, we also replicate the processing method in the original work \cite{2018Improving} to extract the train dataset, the validation dataset and the test dataset. As a result, we obtain 50,400 examples for the train dataset, 13,248 for the validation dataset and 13,216 for the test dataset.


For the Java dataset, the first summary sentence of the Javadoc annotations is usually used as the comment, which describes the functionality of the Java method. To be consistent with the original work~\cite{Hu2018Summarizing}, we reuse these extracted comments included in the public dataset. For the Python dataset, we use the comment provided by the source code. Data instances of these datasets are in the form of $\langle p,comment\rangle$ pair, where $p$ is the source code snippet and  $comment$ is the reference comment. We pre-process the dataset by the Javalang \textsuperscript{\ref {javalang}} parser for the Java dataset and the ast\textsuperscript{\ref {ast}} library  for the Python dataset, and discard those  syntactically incorrect programs. Finally, we follow the  processing steps \cite{2020A} which splits camelCase and snake\_case tokens into their corresponding sub-tokens.

\smallskip
\noindent\textbf{Victim models.}
The victim models (i.e., the target models under adversarial attacks) in our experiments are based on LSTM, Transformer, GNN, a dual model (CSCG), and a retrieval-based neural source code summarization model named Rencos.
\begin{itemize}
\item {LSTM-based seq2seq model.} A LSTM-based seq2seq model \cite{2020A} contains 2-layers BiLSTM for encoder and decoder with attention mechanism, encoding the source code to an intermediate representation and translating it to natural language, i.e., comment.
\item{Transformer-based seq2seq model.} Ahmad et al. \cite{2020A} designed the Transformer-based seq2seq model  for code comment generation by introducing multi-head attention as encoder and decoder. To the  best of our knowledge, this model represents the state-of-the-art result  on the Java dataset. 
\item{GNN-based seq2seq model.} LeClair et al. \cite{leclair2020codegnn} employed two encoders, one is the GNN-based encoder to model structural information and the other is the GRU-based encoder to model textual information and GRU-based decoder to generate natural language comment. 
\item{CSCG Dual model.} Wei et al. \cite{NIPS2019_8883} designed a dual learning framework to train a code summary i.e comment generation and a code generation model simultaneously using the LSTM-based seq2seq model.
\item{Retrieval-based model (Rencos).} Zhang et al. \cite{rencos} 
	leverage both neural and retrieval-based techniques to enhance the neural model with the most similar code snippets at the syntax-level and the semantics-level.
\end{itemize}
We largely follow the settings of the respective original work; in particular, the hyperparameters of the victim models are listed in Table~\ref{Table:2}. All models were trained and evaluated on a server running Ubuntu 20.04 LTS OS with 2 Intel Xeon 4216 2.10GHz Silver CPUs , and 4 RTX2080Ti GPUs.

\begin{table}[h]
	\begin{center}
		\caption{Hyperparameters in our experiments}
		\setlength{\tabcolsep}{4mm}{
			\begin{tabular}{|c|c|c|c|c|c|}
				\hline
				
				\textbf{Hyperparameters} &\textbf{ LSTM} & \textbf{Transformer} &\textbf{ GNN} & \textbf{CSCG} &\textbf{Rencos}\\ \hline\hline
				n\_layers &2 & 6 & 1 & 3 &1 \\ 
				\hline
				n\_head & -- & 8 & -- & -- & --\\
				\hline 
				d\_k, d\_v & --  & 64 & -- & -- & -- \\ 
				\hline
				d\_ff & -- & 2048 & -- & -- &  --\\ 
				\hline
				embed\_size & 512 & 512 & 256 & 512& 256\\ 
				\hline
				hidden\_size & 512 & -- & 256 & 512 & 512\\ 
				\hline
				optimizer & adam & adam & adam & adam & adam\\ 
				\hline
				learning rate & 0.002& 0.0001 & 0.001 & 0.002 &0.001 \\ 
				\hline
				batch size & 32 & 32 & 32 & 32 & 32\\ \hline
		\end{tabular}}	
		\label{Table:2} 
	\end{center}
\end{table}

\smallskip
\noindent\textbf{Baseline approaches for adversarial attack.}
Since we are the first to consider adversarial examples for code comment generation tasks, the literature is short of algorithms for direct comparison. To demonstrate the effectiveness of our approach, we adopt two algorithms as the baseline, i.e., the random substitute algorithm and the algorithm based on Metropolis-Hastings sampling \cite{zhang2020generating}. 
\begin{description}
	\item[Random substitution.]
The random substitute algorithm is a na\"ive algorithm where both the substituted identifiers and candidate identifiers are randomly sampled. 

\item[Metropolis-Hastings algorithm.]
The Metropolis-Hastings sampling based algorithm was recently used to generate adversarial examples for attacking source code classifiers \cite{zhang2020generating}. 
Recall that the Metropolis-Hastings algorithm is a classical Markov Chain Monte Carlo sampling approach, which can generate desirable examples given the targeted stationary distribution and the transition proposal. We adapt the algorithm \cite{Chib1995Understanding} to our code comment generation task. 
\end{description}

In general, we want the adversarial examples to be as close to the original example as possible. For this purpose, we set $max$, the maximum number of identifiers that can be substituted. 
(In the current experiments we set $max$ to be 2 or 3.) 

%
 
\smallskip
\noindent \textbf{Metrics.} As the generated comments are in natural language, we adopt the standard metrics from neural machine translation, i.e., BLEU, METEOR, ROUGE-L to measure the quality of the generated comments.
The lower these values are after attack, the higher the degradation of comment generation models is, i.e., the less robust these models are.
%
%
%
%
Moreover, we introduce three additional metrics to evaluate the performance of different adversarial example generation algorithms. 
\begin{itemize}
	\item Relative degradation. We follow Michel et al.'s work~\cite{2019On} to measure the  (relative) degradation of the model under attack. Formally,
	\[  
	r_d=\frac{\textrm{BLEU}(y, \mbox{refs})-\textrm{BLEU}(y', \mbox{refs})}{ \textrm{BLEU}(y, \mbox{refs})} ,
	\]
	where refs denotes the reference comment, $y$ is the original output, and $y'$ is output of the perturbed program. 
	
	\item Valid rate, which is defined as the percentage of generated adversarial examples which can pass the compilation. Formally, 
		\[
	v_r=\frac{Count_{valid}}{Count_{all}}.  
	\] 
	This metric is used to assess the quality of the generated adversarial examples, as well as the efficiency of the generation process.

	\item Success rate, which is defined as the product of the relative degradation and the valid rate, providing a comprehensive indicator of attack efficiency and example quality. Formally, 
		\[
	s_r=r_d*v_r. 
	\]
	Essentially a higher success rate indicates the corresponding method can generate valid adversarial examples with better attack capability, hence 
	entails a more effective attack method.  

\end{itemize}

\subsection{Research questions and results}\label{Results}
In our experiments, we primarily investigate the following four research questions (RQs).
\begin{description}
	\item[RQ1.] Are existing code comment generation models vulnerable to our adversarial attacks?
	\item[RQ2.] How effective is our adversarial attack method, i.e., how successful can it achieve to attack code comment generation models over the baseline methods?
	\item[RQ3.] Do adversarial samples generated by our adversarial attack method have better transferability than the baseline methods?
	\item[RQ4.] How efficient is the masked training method in improving robustness?
\end{description}

\medskip
\noindent \textbf{RQ1. Are existing code comment generation models vulnerable to our adversarial attacks?} 

To answer this research question, we generate adversarial examples on the test dataset using  \toolname\ to attack four different models. The performance of different models before and after the attack is listed in Table~\ref{Table:3}.

\begin{table}[h]
	\color{black}{
		\begin{center}
			\caption{Results of adversarial attack on different models \small{(`$max$' means the maximum substitution identifier number used in different methods; `original' is the  result on the clean test set.)} }
			\begin{adjustbox}{width=\columnwidth,center}
			{
				\begin{tabular}{|c|c|c|c|c|c|c|c|}
					\hline
					\multicolumn{2}{|c|}{\multirow{2}{*}{}}                                                                 & \multicolumn{3}{c|}{\textbf{Java Dataset}}             & \multicolumn{3}{c|}{\textbf{Python Dataset}}            \\ \cline{3-8} 
					\multicolumn{2}{|c|}{}                                                                                  & \textbf{BLEU}           &\textbf{ METEOR}        & \textbf{ROUGE-L}        & \textbf{BLEU }          & \textbf{METEOR}        & \textbf{ROUGE-L }       \\ \hline\hline
					\multirow{3}{*}{LSTM}       & original                                                        & 35.47          & 19.72         & 47.57          & 30.83          & 17.06         & 41.77          \\ \cline{2-8} 
					& \begin{tabular}[c]{@{}c@{}}attack\\ max=2\end{tabular} &13.08 & 6.83& 21.75 &18.30 & 8.62 & 27.24 \\ \cline{2-8} 
					& \begin{tabular}[c]{@{}c@{}}attack\\ max=3\end{tabular} & 13.07 & 6.75 & 21.52 & 17.92 & 8.27 &26.64\\ \hline\hline
					\multirow{3}{*}{Transformer} & original                                                        & 44.58          & 26.43         & 54.76          & 33.15          & 18.96         & 44.50          \\ \cline{2-8} 
					& \begin{tabular}[c]{@{}c@{}}attack\\ max=2\end{tabular} & 13.23 &8.08 &22.79 & 18.87 & 8.99 &27.91\\ \cline{2-8} 
					& \begin{tabular}[c]{@{}c@{}}attack\\ max=3\end{tabular} & 13.14 &7.90 & 22.42 & 18.54 & 8.57& 27.29 \\ \hline\hline
					\multirow{3}{*}{GNN}        & original                                                        & 39.41          & 23.32         & 46.65          & 31.24              & 15.77            & 38.28              \\ \cline{2-8} 
					& \begin{tabular}[c]{@{}c@{}}attack\\ max=2\end{tabular} &16.44 &7.77 &21.36 & 19.38  & 7.36   & 24.07    \\ \cline{2-8} 
					& \begin{tabular}[c]{@{}c@{}}attack\\ max=3\end{tabular} &16.14  & 7.42 & 20.55 &  18.65   & 6.59    & 22.72  \\ \hline\hline
					\multirow{3}{*}{CSCG}        & original                                                       & 42.39          & 25.77         & 53.61          & 30.82          & 17.67         & 48.14          \\ \cline{2-8} 
					& \begin{tabular}[c]{@{}c@{}}attack\\ max=2\end{tabular} & 8.85 & 5.10& 23.50 & 11.90 & 6.23 & 32.94 \\ \cline{2-8} 
					& \begin{tabular}[c]{@{}c@{}}attack\\ max=3\end{tabular} & 8.95 & 5.02& 23.44 & 12.08 & 6.15 &32.83\\ \hline
					
					\multirow{3}{*}{Rencos} &original  &44.0  & 25.73 & 54.02 &  33.34 & 18.65  & 43.37         \\ \cline{2-8} 
					& \begin{tabular}[c]{@{}c@{}}attack\\ max=2\end{tabular} & 40.68 & 23.09 & 49.17 & 31.02& 15.78& 38.84 \\ \cline{2-8} 
					& \begin{tabular}[c]{@{}c@{}}attack\\ max=3\end{tabular} & 40.23& 22.69& 48.46 & 30.55 & 15.19 & 37.90 \\ \hline
			\end{tabular}}
		\end{adjustbox}
			\label{Table:3}
		\end{center}
	}
\end{table}

We can observe that all code comment generation models are vulnerable to our adversarial attack. When modifying maximum 2 or 3 identifiers in the source code,  the performance of models degrades sharply  in general, although the impact of the adversarial attack differs among these models. The CSCG Dual model has the worst performance on the two datasets. When we test the CSCG Dual model with $max=2$, the BLEU value is only 8.85 on the Java dataset and 11.90 on the Python dataset which means that the model's output is almost meaningless and of little help to program comprehension. The retrieval-based model Rencos performs better under the adversarial attack.  From Table~\ref{Table:3} we can also see that models which are with the structural information (GNN-based seq2seq model) or with the help of most similar code snippets (Rencos) are more robust than models with only contextual information (LSTM-based, Transformer-based and CSCG Dual models). 



To summarize, existing code comment generation models are of poor robustness under adversarial attacks, especially the seq2seq models with only contextual information. 

\medskip\begin{table}[h]
	\color{black}{
		\begin{center}
			\caption{Evaluation of different adversarial examples generation algorithms}
			\begin{tabular}{|c|c|c|c|c|c|c|c|}
				\hline
				\multicolumn{2}{|c|}{\multirow{3}{*}{}} & \multicolumn{3}{c|}{\textbf{Java Dataset}}   & \multicolumn{3}{c|}{\textbf{Python Dataset}} \\ \cline{3-8} 
				\multicolumn{2}{|c|}{}                  & \multicolumn{6}{c|}{\textbf{max=2}}                                \\ \cline{3-8} 
				\multicolumn{2}{|c|}{}                 &\textbf{ $r_d$(\%)}      & \textbf{$v_r$(\%)}       & \textbf{$s_r$(\%)}       & \textbf{$r_d$(\%)   }    &\textbf{ $v_r$(\%)}       & \textbf{$s_r$(\%)    }     \\ \hline
				\multirow{3}{*}{LSTM}         & Random   & 31.38 & 30.82& 9.67  & 23.39 & 27.48 & 6.43  \\ \cline{2-8} 
				&MH      & 42.09 & 100   & 42.09 & 40.29 & 100   & 40.29 \\ \cline{2-8} 
				& ACCENT   & 63.12 & 100   & 63.12 & 40.64 & 100   & 40.64 \\ \hline
				\multirow{3}{*}{Transformer}  & Random   & 38.58 & 30.82 & 11.89 & 22.29 & 27.48 & 6.13  \\ \cline{2-8} 
				& MH      & 64.72 & 100   & 64.72 & 41.45 & 100   & 41.45 \\ \cline{2-8} 
				&  ACCENT & 70.32 & 100   & 70.32 & 43.08 & 100   & 43.08\\ \hline
				\multirow{3}{*}{GNN}          & Random   & --       & --       & --       & --       & --       & --       \\ \cline{2-8} 
				& MH      & 57.14 & 100  & 57.14 & 34.41 & 100   & 34.41 \\ \cline{2-8} 
				&  ACCENT  & 58.28 & 100   & 58.28 & 37.80 & 100   & 37.80 \\ \hline
				\multirow{3}{*}{CSCG}         & Random   & 44.44 & 30.82 & 13.69 & 23.56 & 27.48 & 6.47  \\ \cline{2-8} 
				& MH      & 68.25 & 100   & 68.25 & 37.77 & 100   & 37.77 \\ \cline{2-8} 
				& ACCENT & 79.12 & 100   & 79.12 & 61.39 & 100   & 61.39 \\ \hline
				\multirow{3}{*}{Rencos }        
				& Random  & -- & --  & --  &  -- &  --  &  --   \\ \cline{2-8} 
				& MH      & 6.84 & 100   & 6.84 & 7.11 & 100   & 7.11 \\ \cline{2-8} 
				& ACCENT  & 7.55 & 100   & 7.55 & 6.96 & 100   & 6.96 \\ \hline
				\multicolumn{2}{|c|}{\multirow{2}{*}{}} & \multicolumn{6}{c|}{\textbf{max=3} }                               \\ \cline{3-8} 
				\multicolumn{2}{|c|}{}                 & \textbf{$r_d$(\%)   }    & \textbf{$v_r$(\%)  }     & \textbf{$s_r$(\%)   }    & \textbf{$r_d$(\%)    }   &\textbf{ $v_r$(\%)   }    &\textbf{ $s_r$(\%) }        \\ \hline
				\multirow{3}{*}{LSTM}        & Random   & 32.03 & 29.7 & 9.51  & 21.89 & 27.81 & 6.08  \\ \cline{2-8} 
				& MH      & 44.60 & 100   & 44.60 & 38.63 & 100   & 38.63 \\ \cline{2-8} 
				&  ACCENT &63.15 & 100   & 63.15 & 41.87& 100   & 41.87 \\ \hline
				\multirow{3}{*}{Transformer}  & Random   & 45.76 & 29.7  & 13.59 & 25.52 & 27.81 & 7.09  \\ \cline{2-8} 
				& MH      & 66.42 & 100  & 66.42 & 42.29 & 100   & 42.29 \\ \cline{2-8} 
				& ACCENT  & 70.52& 100   & 70.52 & 44.07 & 100  & 44.07\\ \hline
				\multirow{3}{*}{GNN}          &Random   & --       & --       & --       & --      & --       & --       \\ \cline{2-8} 
				& MH      & 58.51 & 100   & 58.51 & 36.33 & 100   & 36.33 \\ \cline{2-8} 
				& ACCENT  & 59.05 & 100  & 59.05 & 40.30 & 100   & 40.30 \\ \hline
				\multirow{3}{*}{CSCG}         & Random   & 46.36 & 29.7  & 13.76   & 25.44& 27.81 & 7.07  \\ \cline{2-8} 
				& MH      & 68.88 & 100   & 68.88 & 39.00 & 100  & 39.00 \\ \cline{2-8} 
				& ACCENT & 78.89 & 100   & 78.89 & 60.80 & 100   & 60.80 \\ \hline
				
				\multirow{3}{*}{Rencos}     & Random   & -- &  -- &  --  &-- & -- & -- \\ \cline{2-8} 
				& MH      & 7.61 &100   &7.61 & 8.34 & 100  & 8.34 \\ \cline{2-8} 
				& ACCENT & 8.57 &100   &8.57 &  8.37& 100   &8.37  \\ \hline
			\end{tabular} 
			
			\label{Table:4}
		\end{center}
	}
\end{table}


\begin{table}[h]
	\centering
	\caption{Results of adversarial attack using Random substitution algorithm and MH-based algorithm on different models with $max=2$}
			\begin{adjustbox}{width=\columnwidth,center}
		\begin{tabular}{|c|c|c|c|c|c|c|c|}
			\hline
			\multicolumn{2}{|c|}{\multirow{2}{*}{}} & \multicolumn{3}{c|}{\textbf{Java Dataset}} & \multicolumn{3}{c|}{\textbf{Python Dataset}} \\ \cline{3-8} 
			\multicolumn{2}{|c|}{}                  & \textbf{BLEU}   & \textbf{METEOR} & \textbf{ROUGE-L} & \textbf{BLEU}   & \textbf{METEOR}  & \textbf{ROUGE-L}  \\ \hline
			\multirow{2}{*}{LSTM}         &Random  & 24.34  & 13.45  & 35.43   & 23.62  & 11.69   & 32.94    \\ \cline{2-8} 
			& MH      & 20.54  & 10.81  & 30.3    & 18.41  & 8.77    & 27.85    \\ \hline
			\multirow{2}{*}{Transformer}  & Random & 27.38  & 15.61  & 37.26   & 25.76  & 13.35   & 35.81    \\ \cline{2-8} 
			& MH      & 15.73  & 9.68   & 26.74   & 19.41  & 9.98    & 30.19    \\ \hline
			\multirow{2}{*}{GNN}          & Random  & --      & --      & --       & --      & --       & --        \\ \cline{2-8} 
			& MH      & 16.89  & 8.38   & 22.13  & 20.49  & 8.10    & 25.21    \\ \hline
			\multirow{2}{*}{CSCG}         & Random  & 23.55  & 13.50  & 39.65   & 23.56  & 12.08   & 40.23    \\ \cline{2-8} 
			& MH      & 13.46  & 7.72   & 29.65   & 19.18  & 9.79    & 37.01    \\ \hline
			\multirow{2}{*}{Rencos}         & Random  &  --  &  -- &  --  & -- & -- &  --  \\ \cline{2-8} 
			&MH   & 40.99& 23.44   & 49.84  & 30.97 & 15.48  &  38.27   \\ \hline
	\end{tabular} 
   \end{adjustbox}
	\label{Table:5}
\end{table}
\begin{table}[h]
	\centering
	\caption{Results of adversarial attack using Random substitution algorithm and MH-based algorithm on different models with $max=3$}
		\begin{adjustbox}{width=\columnwidth,center}
		\begin{tabular}{|c|c|c|c|c|c|c|c|}
			\hline
			\multicolumn{2}{|c|}{\multirow{2}{*}{}} & \multicolumn{3}{c|}{\textbf{Java Dataset}} & \multicolumn{3}{c|}{\textbf{Python Dataset}} \\ \cline{3-8} 
			\multicolumn{2}{|c|}{}                  & \textbf{BLEU}   & \textbf{METEOR} & \textbf{ROUGE-L} & \textbf{BLEU}   & \textbf{METEOR}  & \textbf{ROUGE-L}  \\ \hline
			\multirow{2}{*}{LSTM}         & Random  & 24.11  & 13.24  & 35.15   & 24.08  & 12.09   & 33.52    \\ \cline{2-8} 
			& MH      & 19.65  & 10.2   & 28.99   & 18.92  & 9.23    & 28.64    \\ \hline
			\multirow{2}{*}{Transformer}  & Random  & 24.18  & 19.72  & 42.13   & 24.69  & 12.60   & 34.76    \\ \cline{2-8} 
			& MH     & 14.97  & 9.17   & 25.59   & 18.80  & 9.27    & 29.04    \\ \hline
			\multirow{2}{*}{GNN}          & Random  & --      & --      & --       & --      & --       & --        \\ \cline{2-8} 
			& MH       & 16.35  & 7.70   & 20.95  & 19.89  & 7.37    & 24.17    \\ \hline
			\multirow{2}{*}{CSCG}         & Random  & 22.74  & 12.96  & 38.99   & 22.98  & 11.54   & 39.49    \\ \cline{2-8} 
			& MH      & 13.19  & 7.28   & 28.83   & 18.80  & 9.35    & 36.30    \\ \hline
			\multirow{2}{*}{Rencos}    &  Random  & --  &  --   &  --   & --  &  --  &  --   \\ \cline{2-8} 
			& MH  & 40.65 & 23.19  & 49.42 & 30.56 & 14.99   & 37.57 \\ \hline
	\end{tabular}
\end{adjustbox}
	\label{Table:6}
\end{table}

\noindent\textbf{RQ2. How effective is our adversarial attack method, i.e., how successful can it achieve to attack code comment generation models over the baseline methods?} 

For this research question, we analyze the effectiveness of different algorithms on the five models across two datasets with $max=2$ and $max=3$. The results are given in Table~\ref{Table:4}, Table~\ref{Table:5}, and Table~\ref{Table:6}.
As the input of the GNN-based model and the retrieval-based model need to be compiled to generate AST, only a small part of the samples generated by the random substitution algorithm are valid samples (i.e., can be compiled), hence only the MH-based method and the {\toolname} attack method are compared in the two models. In other models, baseline algorithms contain random substitution algorithm, MH-based algorithm, and our {\toolname} attack method. Taking the original models' BLEU as the standard performance metric, the {\toolname} attack method can reduce the performance by 63.12\% for LSTM, 70.32\% for Transformer, 58.28\% for GNN, 79.12\% for CSCG and 7.55\% for Rencos on the Java dataset, and 40.64\% for LSTM, 43.08\% for Transformer, 37.80\% for GNN, 37.77\% for CSCG and 6.96\% for Rencos with $max=2$, which are considerably better than the baselines. When $max$ is 3, our attacking method can degrade the model performance even further. The effectiveness of the adversarial samples generated by the random substitution algorithm is extremely low, while the adversarial samples generated by the {\toolname} attack method and the MH-based algorithm can guarantee 100\%
effectiveness. That means they are all correct code snippets in grammar, and the {\toolname} attack method can achieve a higher success rate.

To further investigate the effectiveness of our approach, 
we apply the Mann-Whitney U test. Particularly, we compare {\toolname} attack method with MH, and test whether the effectiveness of the former is significantly better than the latter. We focus on the $r_d$ values of {\toolname} and MH. 
For each run, we randomly sample 100 Java code snippets, and 100 Python code snippets from the two datasets, and calculate the average $r_d$ values of the generated comments by the two attack methods on the five base models as outcomes. The experiment is repeated 5 times with max=2 and max=3 respectively. As a result, there are in total of 20 experiments (i.e., max=2 or 3 for five based models and for Java and Python datasets). For each one of them, we obtain two samples of size 5. 
In the hypothesis test, we follow the convention to set $\alpha=0.05$. For the Mann-Whitney U test, 
a majority of p-values (15 out of 20) are less than 0.05 (typically 0.005), which indicates that the improvements are statistically significant at the confidence level of 95\%.\footnote{The details of the samples can be retrieved in our replication package.}
To conclude, the adversarial samples generated by {\toolname} are effective, and our attack method is superior to the baseline methods.

\begin{table}[h]
	\color{black}{
		\centering
		\caption{BLEU scores of different algorithms for transferability on Java dataset ($max=2$).}
		\begin{tabular}{|c|c|c|c|c|c|}
			\hline
			\rowcolor{gray!20}
			\multicolumn{2}{|c|}{}                & CSCG  & LSTM  & Transformer & Rencos      \\ \hline
			\multirow{2}{*}{\begin{tabular}[c]{@{}c@{}}Adversairal examples\\ generated for GNN\end{tabular}}         & MH     & 17.69 & 18.91 & 22.26       & 43.37         \\
			& ACCENT & 16.22 & 18.10 & 21.59       & 43.39          \\ 
			\hline
			\rowcolor{gray!20}
			&        & GNN   & LSTM  & Transformer & Rencos      \\ \hline
			\multirow{2}{*}{\begin{tabular}[c]{@{}c@{}}Adversairal examples\\  generated for CSCG\end{tabular}}       & MH     & 15.78 & 17.37 & 20.50       & 43.43         \\
			& ACCENT & 15.31 & 15.84 & 19.06       & 43.32           \\ \hline
			\rowcolor{gray!20}
			&        & GNN   & CSCG  & Transformer & Rencos      \\ \hline
			\multirow{2}{*}{\begin{tabular}[c]{@{}c@{}}Adversairal examples\\  generated for LSTM\end{tabular}}       & MH     & 19.39 & 17.91 & 23.89       & 43.44           \\
			& ACCENT & 16.05 & 14.48 & 19.10       & 43.39          \\ \hline
			\rowcolor{gray!20}
			&        & GNN   & CSCG  & LSTM        & Rencos      \\ \hline
			\multirow{2}{*}{\begin{tabular}[c]{@{}c@{}}Adversairal examples\\ generated for Transformer\end{tabular}} & MH     & 16.11 & 15.07 & 16.59       & 43.43        \\
			& ACCENT & 15.92 & 14.42 & 15.83       & 43.35          \\ \hline
			\rowcolor{gray!20}
			&        & GNN   & CSCG  & LSTM        & Transformer \\ \hline
			\multirow{2}{*}{\begin{tabular}[c]{@{}c@{}}Adversairal examples \\ generated for Rencos\end{tabular}}     & MH     & 17.69 & 19.95 & 21.28      & 24.46     \\
			& ACCENT & 17.50 & 18.85 & 20.26       & 24.12       \\ \hline
		\end{tabular}
		\label{java2}
	}
\end{table}
\begin{table}[h]
	\centering

		\caption{BLEU scores of different algorithms for transferability on Java dataset ($max=3$).}
		\begin{tabular}{|c|c|c|c|c|c|}
			\hline
			\rowcolor{gray!20}
			\multicolumn{2}{|c|}{}                            & CSCG  & LSTM  & Transformer & Rencos      \\ \hline
			\multirow{2}{*}{\begin{tabular}[c]{@{}c@{}}Adversairal examples\\ generated for GNN\end{tabular}}         & MH     & 16.33 & 18.39 & 20.76       & 43.37          \\
			& ACCENT & 15.30 & 17.38 & 20.45       & 43.45          \\ \hline
			\rowcolor{gray!20}
			&        & GNN   & LSTM  & Transformer & Rencos      \\ \hline
			\multirow{2}{*}{\begin{tabular}[c]{@{}c@{}}Adversairal examples\\  generated for CSCG\end{tabular}}       & MH     & 15.58 & 16.75 & 29.12       & 43.36         \\
			& ACCENT & 15.25 & 15.35 & 18.27       & 43.27          \\ \hline
			\rowcolor{gray!20}
			&        & GNN   & CSCG  & Transformer & Rencos      \\ \hline
			\multirow{2}{*}{\begin{tabular}[c]{@{}c@{}}Adversairal examples\\  generated for LSTM\end{tabular}}       & MH     & 19.68 & 19.52 & 22.28       &43.42         \\
			& ACCENT & 15.86 & 13.93 & 18.27       & 43.40           \\ \hline
			\rowcolor{gray!20}
			&        & GNN   & CSCG  & LSTM        & Rencos      \\ \hline
			\multirow{2}{*}{\begin{tabular}[c]{@{}c@{}}Adversairal examples\\ generated for Transformer\end{tabular}} & MH     & 15.81 & 14.71 & 16.00       & 43.40          \\
			& ACCENT & 15.87 & 14.10 & 15.32       & 43.37         \\ \hline
			\rowcolor{gray!20}
			&        & GNN   & CSCG  & LSTM        & Transformer \\ \hline
			\multirow{2}{*}{\begin{tabular}[c]{@{}c@{}}Adversairal examples \\ generated for Rencos\end{tabular}}     & MH     & 16.82 & 17.87 & 19.26       & 22.24       \\
			& ACCENT & 15.19 & 16.71 & 18.48       & 21.70       \\ \hline
		\end{tabular}
		\label{java3}

\end{table}

\begin{table}[h]
	\centering
	\color{black}{
		\caption{BLEU scores of different algorithms for transferability on Python dataset ($max=2$).}
		\begin{tabular}{|c|c|c|c|c|c|}
			\hline
			\rowcolor{gray!20}
			\multicolumn{2}{|c|}{}                                     & CSCG  & LSTM  & Transformer & Rencos      \\ \hline
			\multirow{2}{*}{\begin{tabular}[c]{@{}c@{}}Adversairal examples\\ generated for GNN\end{tabular}}         & MH     & 22.65 & 22.63 & 24.17       & 32.76           \\
			& ACCENT & 21.49 & 22.79 & 24.32       & 32.54           \\ \hline
			\rowcolor{gray!20}
			&        & GNN   & LSTM  & Transformer & Rencos      \\ \hline
			\multirow{2}{*}{\begin{tabular}[c]{@{}c@{}}Adversairal examples\\  generated for CSCG\end{tabular}}       & MH     & 20.15 & 22.38 & 23.49       & 32.88           \\
			& ACCENT & 16.92 & 22.96 & 21.35       & 30.44          \\ \hline
			\rowcolor{gray!20}
			&        & GNN   & CSCG  & Transformer & Rencos      \\ \hline
			\multirow{2}{*}{\begin{tabular}[c]{@{}c@{}}Adversairal examples\\  generated for LSTM\end{tabular}}       & MH     & 20.18 & 21.74 & 22.81       & 32.75        \\
			& ACCENT & 19.04 & 20.17 & 22.36       & 32.28           \\ \hline
			\rowcolor{gray!20}
			&        & GNN   & CSCG  & LSTM        & Rencos      \\ \hline
			\multirow{2}{*}{\begin{tabular}[c]{@{}c@{}}Adversairal examples\\ generated for Transformer\end{tabular}} & MH     & 20.27 & 21.99 & 21.39       & 32.82           \\
			& ACCENT & 18.89 & 20.10 & 20.91       & 32.28           \\ \hline
			\rowcolor{gray!20}
			&        & GNN   & CSCG  & LSTM        & Transformer \\ \hline
			\multirow{2}{*}{\begin{tabular}[c]{@{}c@{}}Adversairal examples \\ generated for Rencos\end{tabular}}     & MH     & 20.58 & 22.67 & 22.85       &    23.95         \\
			& ACCENT & 19.69 & 22.18 & 22.92       & 24.71       \\ \hline
		\end{tabular}
		\label{python2}
	}
	
\end{table}
\begin{table}[h]
	\centering
	\color{black}{
		\caption{BLEU scores of different algorithms for transferability on Python dataset ($max=3$).}
		\begin{tabular}{|c|c|c|c|c|c|}
			\hline
			\rowcolor{gray!20}
			\multicolumn{2}{|c|}{}                               & CSCG  & LSTM  & Transformer & Rencos      \\ \hline
			\multirow{2}{*}{\begin{tabular}[c]{@{}c@{}}Adversairal examples\\ generated for GNN\end{tabular}}         & MH     & 21.86 & 21.81 & 23.06       & 32.64          \\
			& ACCENT & 20.46 & 21.66 & 22.90       & 32.46         \\ \hline
			\rowcolor{gray!20}
			&        & GNN   & LSTM  & Transformer & Rencos      \\ \hline
			\multirow{2}{*}{\begin{tabular}[c]{@{}c@{}}Adversairal examples\\  generated for CSCG\end{tabular}}       & MH     & 19.88 & 21.49 & 22.60       & 32.79          \\
			& ACCENT & 16.61 & 20.57 & 21.94       &30.46         \\ \hline
			\rowcolor{gray!20}
			&        & GNN   & CSCG  & Transformer & Rencos      \\ \hline
			\multirow{2}{*}{\begin{tabular}[c]{@{}c@{}}Adversairal examples\\  generated for LSTM\end{tabular}}       & MH     & 20.06 & 19.78 & 21.80       & 32.68          \\
			& ACCENT & 18.65 & 19.42 & 21.32       & 32.25           \\ \hline
			\rowcolor{gray!20}
			&        & GNN   & CSCG  & LSTM        & Rencos      \\ \hline
			\multirow{2}{*}{\begin{tabular}[c]{@{}c@{}}Adversairal examples\\ generated for Transformer\end{tabular}} & MH     & 20.19 & 19.92 & 21.49       &32.69         \\
			& ACCENT & 18.46 & 19.40 & 20.50       & 32.17         \\ \hline
			\rowcolor{gray!20}
			&        & GNN   & CSCG  & LSTM        & Transformer \\ \hline
			\multirow{2}{*}{\begin{tabular}[c]{@{}c@{}}Adversairal examples \\ generated for Rencos\end{tabular}}     & MH     & 19.81 & 21.95 & 21.92       & 22.98       \\
			& ACCENT & 18.98 & 21.05 & 21.66       & 23.07       \\ \hline
		\end{tabular}
		\label{python3}
	}
\end{table}

\begin{figure}[htbp]
	\centering
	\subfigure[Java Dataset with $max=2$.]{
		\includegraphics[width=5.5cm]{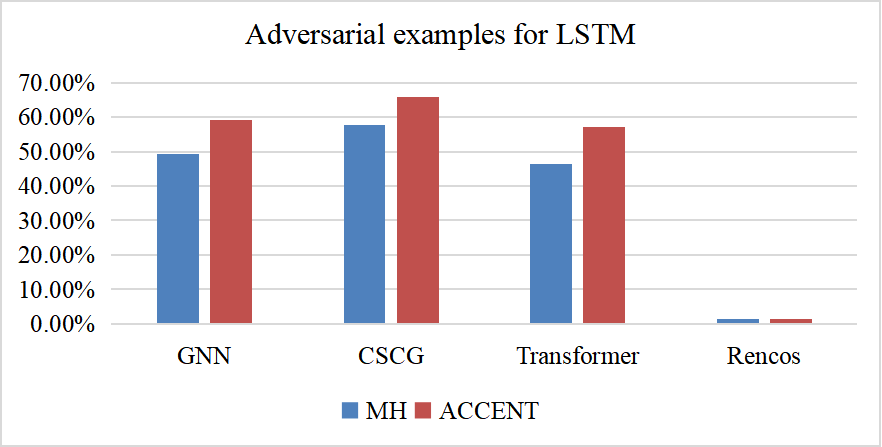}
	}
	\quad
	\subfigure[Python Dataset with $max=2$.]{
		\includegraphics[width=5.5cm]{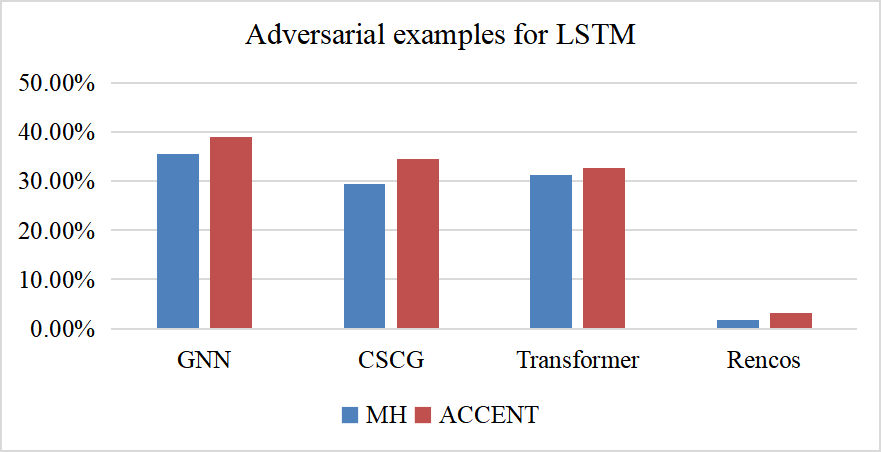}
	}
	\quad
	\subfigure[Java Dataset with $max=3$.]{
		\includegraphics[width=5.5cm]{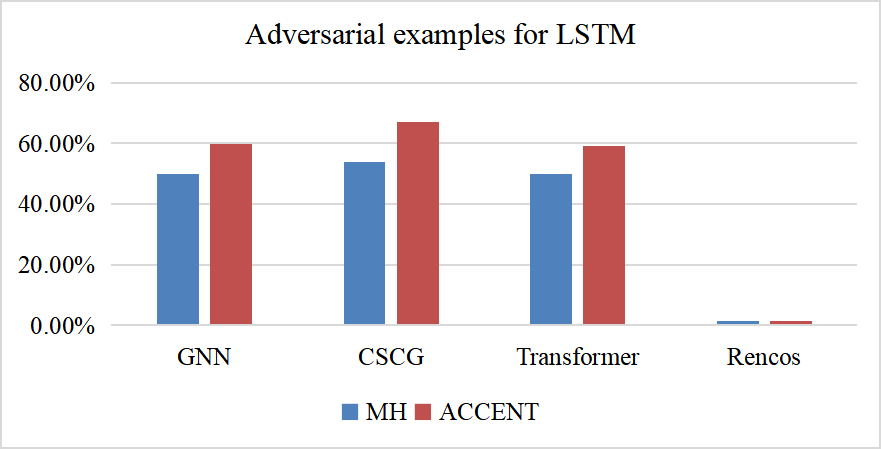}
	}
	\quad
	\subfigure[Python Dataset with $max=3$.]{
		\includegraphics[width=5.5cm]{figs/lstm2p.png}
	}
	\caption{The transferability of adversarial examples generated by different algorithms on LSTM model: the values $r_d$ are tested by attacking GNN, CSCG, Transformer and Rencos model.}
\label{fig:performrack-lstm}
\end{figure}
\begin{figure}[htbp]
	\centering
	\subfigure[Java Dataset with $max=2$.]{
		\includegraphics[width=5.5cm]{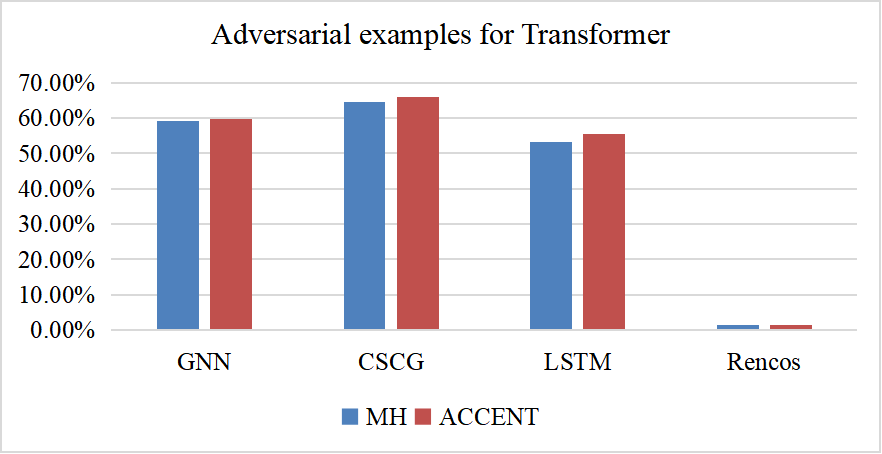}
	}
	\quad
	\subfigure[Python Dataset with $max=2$.]{
		\includegraphics[width=5.5cm]{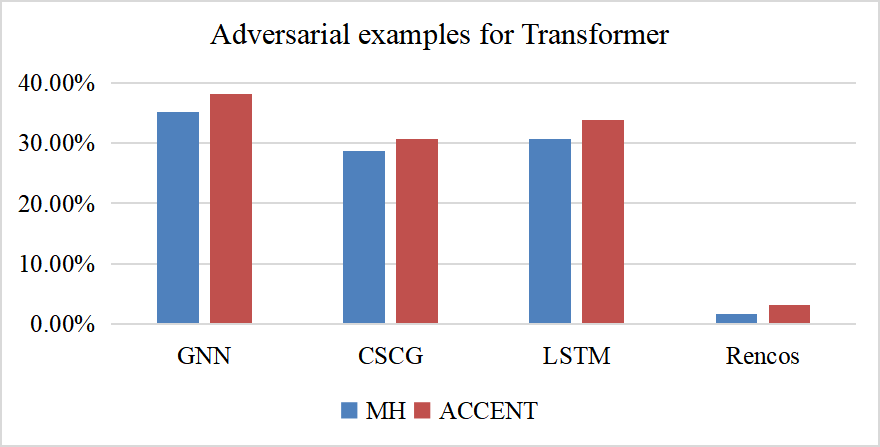}
	}
	\quad
	\subfigure[Java Dataset with $max=3$.]{
		\includegraphics[width=5.5cm]{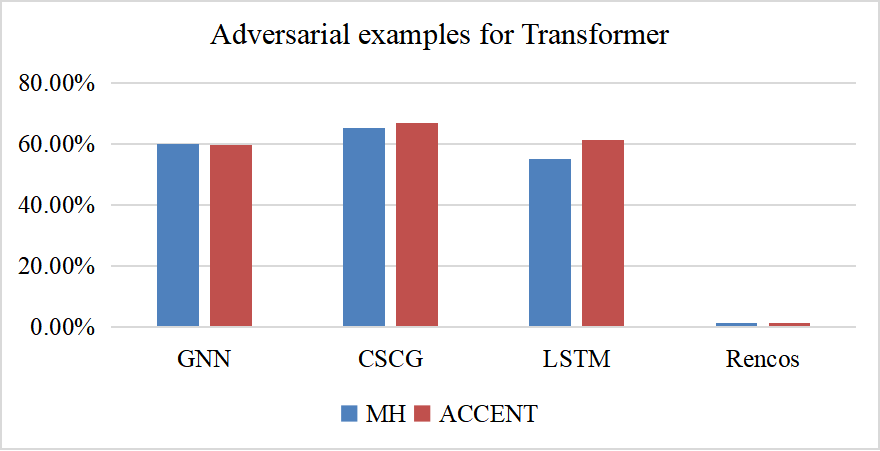}
	}
	\quad
	\subfigure[Python Dataset with $max=3$.]{
		\includegraphics[width=5.5cm]{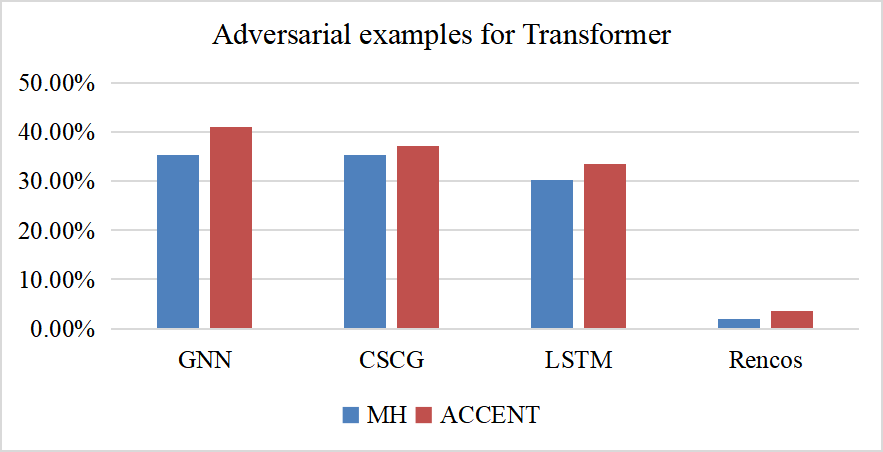}
	}
	\caption{The transferability of adversarial examples generated by different algorithms on Transformer model: the values $r_d$ are tested by attacking GNN, CSCG, LSTM and Rencos model.}
	\label{fig:performrack-trans}
\end{figure}

\begin{figure}[htbp]
	\centering
	\subfigure[Java Dataset with $max=2$.]{
		\includegraphics[width=5.5cm]{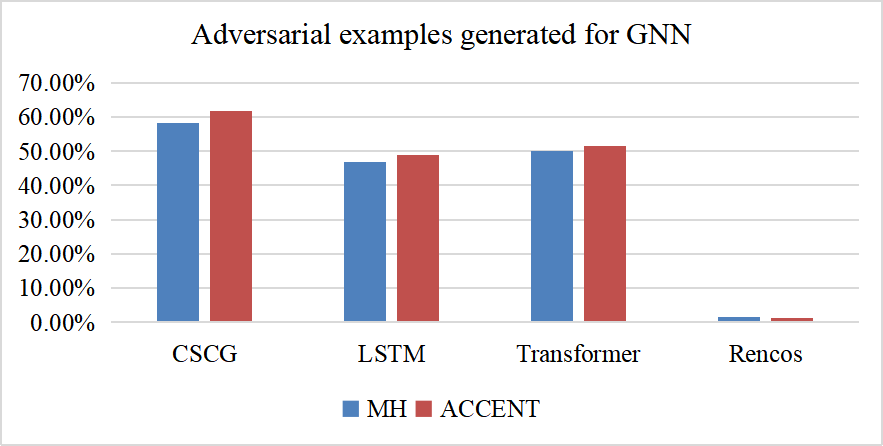}
	}
	\quad
	\subfigure[Python Dataset with $max=2$.]{
		\includegraphics[width=5.5cm]{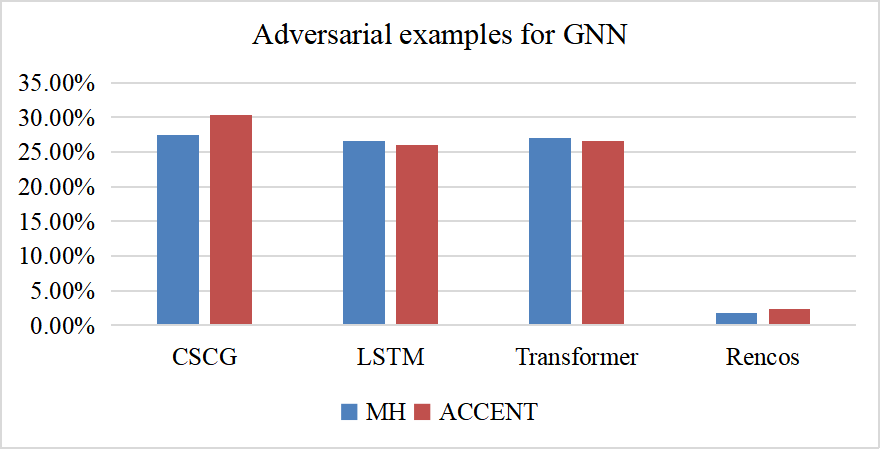}
	}
	\quad
	\subfigure[Java Dataset with $max=3$.]{
		\includegraphics[width=5.5cm]{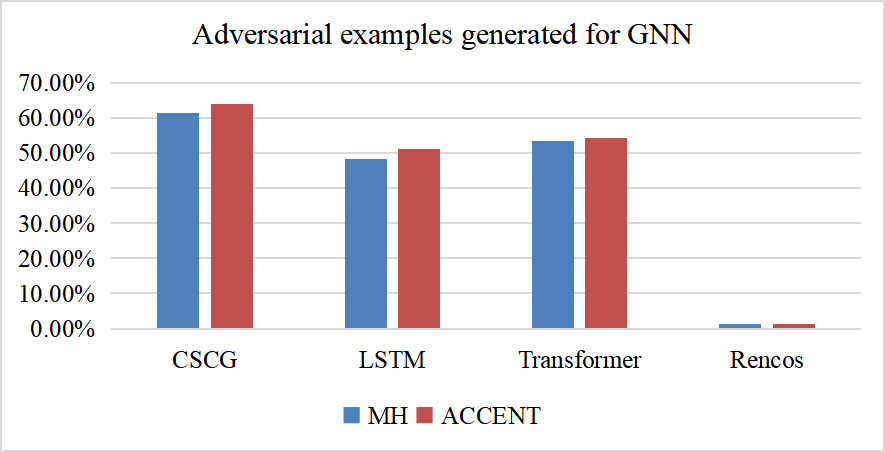}
	}
	\quad
	\subfigure[Python Dataset with $max=3$.]{
		\includegraphics[width=5.5cm]{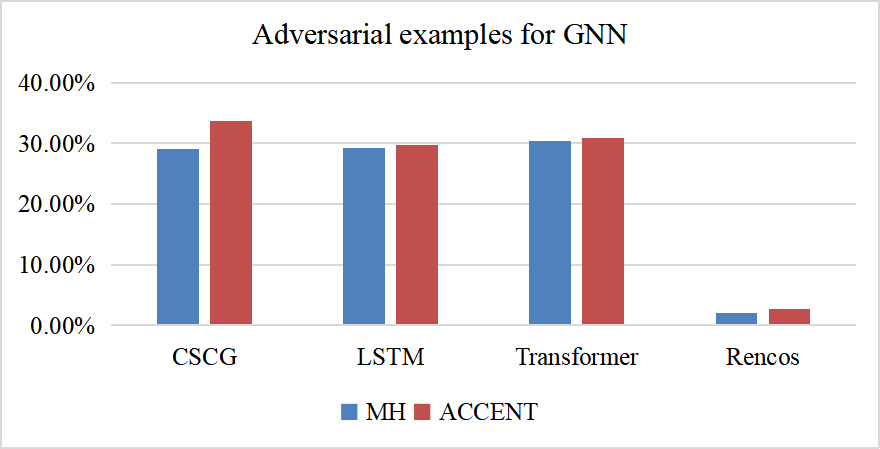}
	}
	\caption{The transferability of adversarial examples generated by different algorithms on GNN model: the values $r_d$ are tested by attacking CSCG, LSTM, Transformer and Rencos model.}
	\label{fig:performrack-gnn}
\end{figure}

\begin{figure}[htbp]
	\centering
	\subfigure[Java Dataset with $max=2$.]{
		\includegraphics[width=5.5cm]{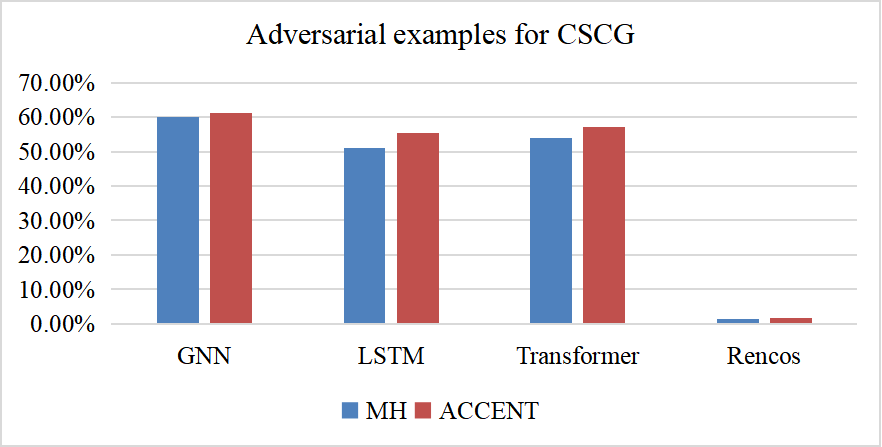}
	}
	\quad
	\subfigure[Python Dataset with $max=2$.]{
		\includegraphics[width=5.5cm]{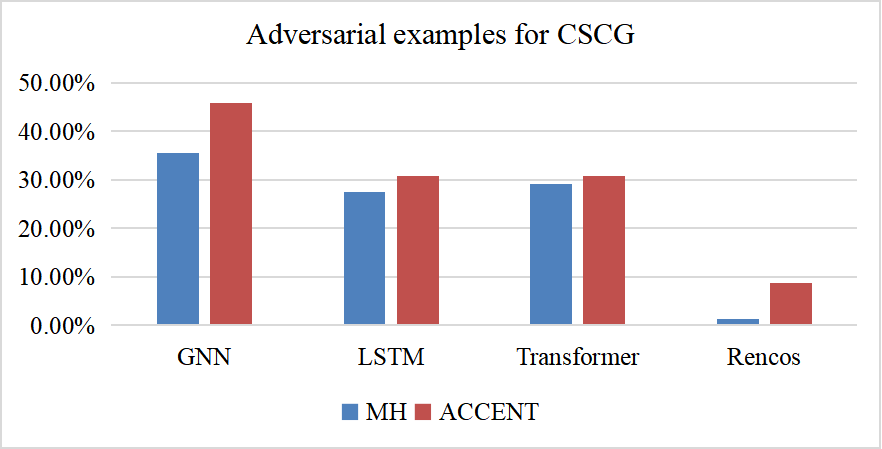}
	}
	\quad
	\subfigure[Java Dataset with $max=3$.]{
		\includegraphics[width=5.5cm]{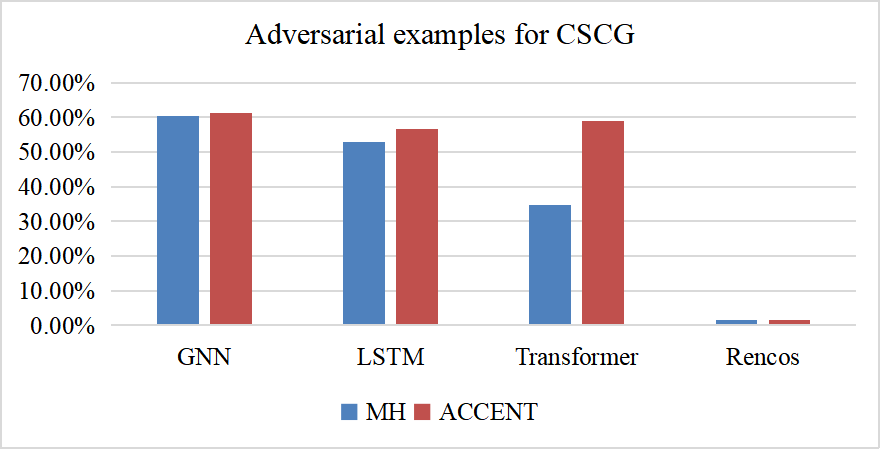}
	}
	\quad
	\subfigure[Python Dataset with $max=3$.]{
		\includegraphics[width=5.5cm]{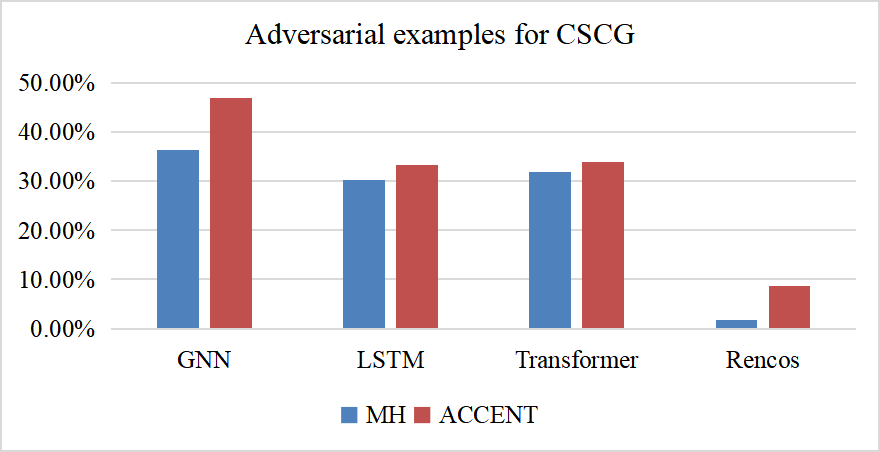}
	}
	\caption{The transferability of adversarial examples generated by different algorithms on CSCG model: the values $r_d$ are tested by attacking GNN, LSTM, Transformer and Rencos model.}
	\label{fig:performrack-cscg}
\end{figure}

\begin{figure}[htbp]
	\centering
	\subfigure[Java Dataset with $max=2$.]{
		\includegraphics[width=5.5cm]{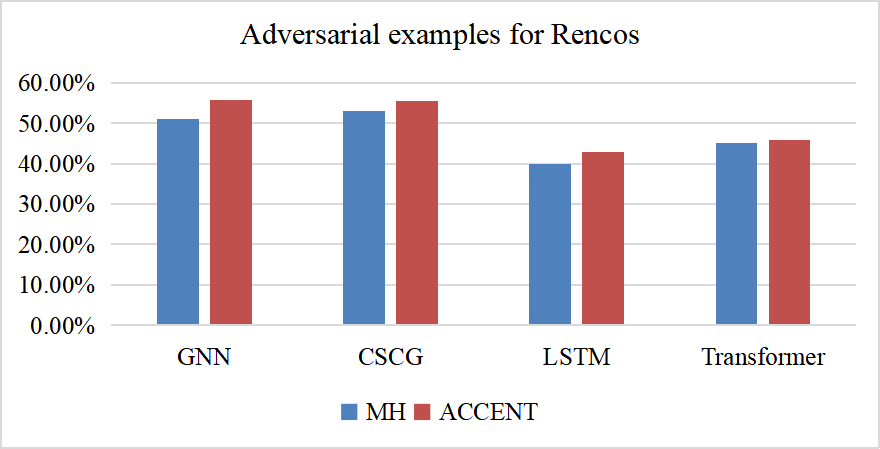}
	}
	\quad
	\subfigure[Python Dataset with $max=2$.]{
		\includegraphics[width=5.5cm]{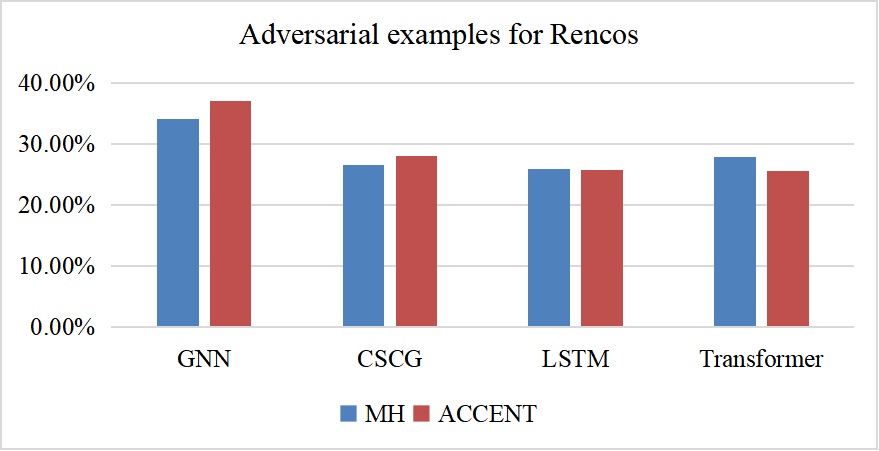}
	}
	\quad
	\subfigure[Java Dataset with $max=3$.]{
		\includegraphics[width=5.5cm]{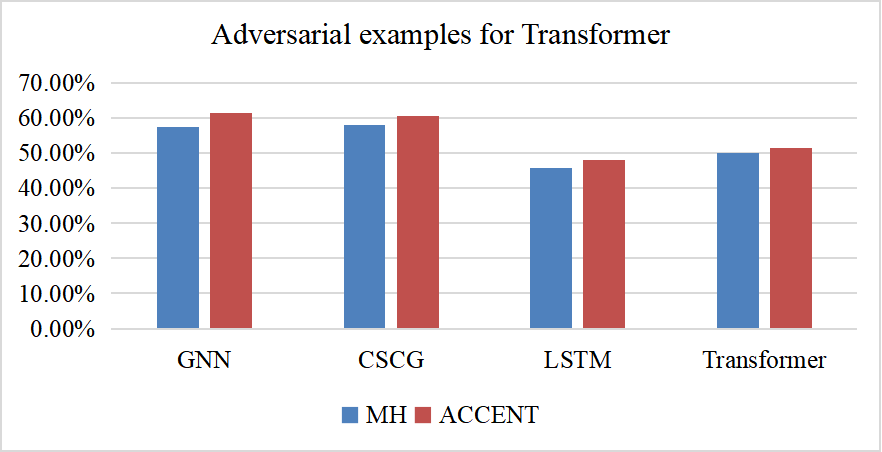}
	}
	\quad
	\subfigure[Python Dataset with $max=3$.]{
		\includegraphics[width=5.5cm]{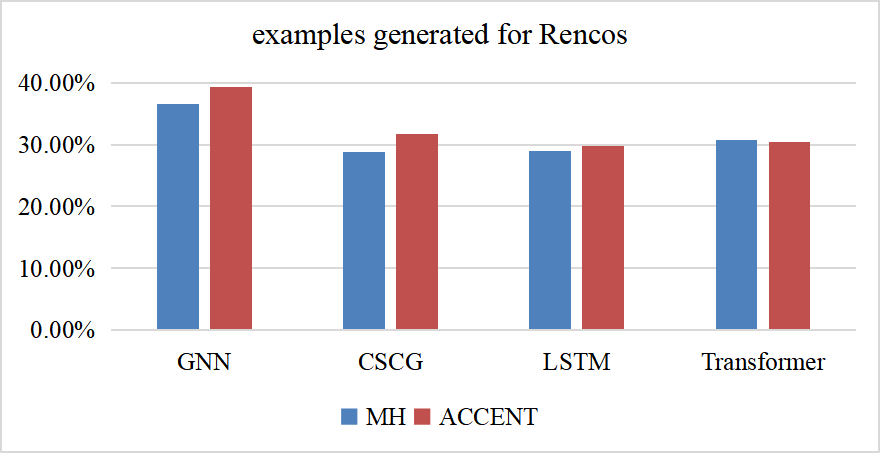}
	}
	\caption{The transferability of adversarial examples generated by different algorithms on Rencos model: the values $r_d$ are tested by attacking GNN, LSTM CSCG and Transformer model.}
	\label{fig:performrack-ir}
\end{figure}


\smallskip
\noindent \textbf{RQ3. Do adversarial samples generated by our adversarial attack method have better transferability than the  baseline methods?}

Adversarial example generated for a certain model is considered to be transferable if it can successfully attack other DNN models. To answer this research question, we tested the transferability of the adversarial examples generated by our {\toolname} attack method and compared them with the MH-based algorithm. The experiment uses a cross-testing method, that is, among the five models, we use the adversarial samples generated from one model to attack the other four models. For example, the adversarial examples generated from the Transformer-based model are used to attack the LSTM-based, GNN-based, CSCGDual models and Rencos. 
The BLEU scores on the Java and the Python datasets are shown in Figure \ref{fig:performrack-lstm}--\ref{fig:performrack-cscg} and Table \ref{java2}--\ref{python3}.

It can be observed from Figure~\ref{fig:performrack-lstm}--Figure~\ref{fig:performrack-ir}  that, except for the Rencos model, the $r_d$ values of the other four models after attacks are decreased by 50\% for the Java dataset, and 37\% for the Python dataset, which means that the performance of the model has dropped greatly, that is, the adversarial samples generated by our {\toolname} attack method can be successfully transferred to other models. At the same time, we can see that, compared with the MH-based algorithm, the adversarial samples generated by the  {\toolname} attack method have better transferability, as the $r_d$ of the {\toolname} attack method is greater than the $r_d$ of the MH-based algorithm on two datasets for all models. This demonstrates that our method can successfully find those identifiers that are important and effective across different models.

\smallskip
\noindent \textbf{RQ4. How efficient is the masked training method in improving robustness?}

We evaluate the effectiveness of our masked training method in improving robustness, which can be evaluated by the changes in performance metrics of DNNs. For each model, we report these metrics (i.e., BLEU, METEOR, and ROGUE-L) over the original test dataset without any perturbations (i.e., `Clean') and the adversarial examples generated by our {\toolname} method with $max=2$ and $max=3$ (i.e., `Adv'). 
We compare the performance of our masked training method with data augmentation, which is a commonly adopted robustness improvement method. In a nutshell, data augmentation improves the robustness by re-training the model with the mixed adversarial dataset which combines the original training dataset with adversarial examples. 
Table~\ref{Table:11} compares the results of our method and the baseline. `Normal' represents the model through the standard training process; `Aug' represents the model trained by data augmentation and `Maksed' represents the model trained by our masked training method.

Improving robustness may sacrifice the accuracy of the models on the clean dataset \cite{belinkov2018synthetic,2018Black,TripletLoss,2018Ensemble}. 
From Table~\ref{Table:11} we can observe that, as the robustness of the model increases, the original accuracy of the model does decrease.
However, our masked training method has less impact on accuracy, where the data augmentation method may suffer from a significant drop. Furthermore, our training method can increase the accuracy of some models on the clean datasets (4 out of 10).
While the accuracy of some models on the clean dataset may slightly decrease, the accuracy on the adversarial examples is improved through our masked training. For example, on the Java dataset, the performance of the Transformer-based model after the masked training has increased to 40.10 and 39.24 on the adversarial examples with $max=2$ and $max=3$ respectively, while the data augmentation method improves the performance to 18.10 and 17.82 respectively.

From Table~\ref{Table:11}, we can conclude  
that the masked training method can significantly boost the robustness across different models at the same time maintain fairly good performance on the test dataset.

\begin{table}[h!]
	\color{black}{
		\begin{center}
			\caption{Results of different methods for improving robustness}
			\begin{adjustbox}{width=\columnwidth,center}
				\begin{tabular}{|c|c|c|c|c|c|c|c|}
					\hline
					\multicolumn{2}{|c|}{\multirow{2}{*}{}}     & \multicolumn{3}{c|}{\textbf{Java Dataset}} & \multicolumn{3}{c|}{\textbf{Python Dataset}} \\ \cline{3-8} 
					\multicolumn{2}{|c|}{}                      & \textbf{BLEU}   & \textbf{METEOR} & \textbf{ROUGE-L} & \textbf{BLEU}   & \textbf{METEOR}  & \textbf{ROUGE-L}    \\ \hline
					\multirow{6}{*}{\small{LSTM}}        & \small{Normal-Clean}    & 35.47     & 19.72     & 47.57      & 30.83    & 17.06      & 41.77       \\ \cline{2-8} 
					& \small{Normal-Adv(max=2)}      & 13.08    & 6.83     & 21.75      & 18.30     & 8.62      & 27.24       \\ \cline{2-8} 
					& \small{Normal-Adv(max=3)}      & 13.07     & 6.75      & 21.52      & 17.92    & 8.27       & 26.64       \\ \cline{2-8} 
					& \small{Aug-Clean}    & 38.14     & 20.96     & 49.22      & 29.36    & 15.58    & 39.71      \\ \cline{2-8} 
					& \small{Aug-Adv(max=2)}      & 22.62    & 11.98    & 32.73      & 23.45    &   10.93   & 31.67      \\ \cline{2-8} 
					& \small{Aug-Adv(max=3)}      & 21.44    & 11.31     & 31.60      & 23.06    & 10.60     & 31.10       \\ \cline{2-8} 
					& \small{Masked-Clean} & 39.60     & 23.24     & 39.60     & 30.64    & 16.70      & 40.54       \\ \cline{2-8} 
					& \small{Masked-Adv(max=2)}   & 31.88   & 18.23    & 41.48    & 27.26    & 13.92      & 36.28\\ \cline{2-8} 
					& \small{Masked-Adv(max=3)}   & 31.31    & 17.84    & 40.85     & 26.81     & 13.56      & 35.72       \\ \hline
					\multirow{6}{*}{\small{Transformer}} & \small{Normal-Clean}    & 44.58     & 26.43     & 54.76      &33.15     & 18.96     & 44.50       \\ \cline{2-8}
					& \small{Normal-Adv(max=2)}     & 13.23    & 8.08    & 22.79      & 18.87     &   8.99  & 27.91      \\ \cline{2-8}
					& \small{Normal-Adv(max=3)}      & 13.14     & 7.90     & 22.42      & 18.54     & 8.57       & 27.29      \\ \cline{2-8} 
					& \small{Aug-Clean}    & 34.14     & 17.36     & 45.77      & 32.97    &  18.76   &   44.02    \\ \cline{2-8} 
					& \small{Aug-Adv(max=2)}      &  18.10  &  9.10   &  27.89    &  24.71   &  12.34   &  33.82    \\ \cline{2-8} 
					& \small{Aug-Adv(max=3)}      &    17.82  &   8.95   &  27.50    &  24.06    &  11.77  & 32.97   \\ \cline{2-8}
					& \small{Masked-Clean} & 44.84     & 27.16   &53.48      & 32.88     & 18.34      & 43.19       \\ \cline{2-8}
					& \small{Masked-Adv(max=2)}      & 40.10  & 24.09 & 48.72    & 28.65   & 14.80     & 37.84     \\ \cline{2-8}
					& \small{Masked-Adv(max=3)}  & 39.24     & 23.46  & 47.88      &28.02     & 14.21     & 37.04    \\ \hline
					\multirow{6}{*}{\small{GNN}}  & Normal-Clean& 39.41 & 23.32  & 46.65    & 31.24     & 15.77      & 38.28       \\ \cline{2-8} 
					& \small{Normal-Adv(max=2)}      & 16.44    & 7.71     & 21.36      & 19.38    &   7.36  & 24.07       \\ \cline{2-8}
					& \small{Normal-Adv(max=3)}      & 16.14     & 7.42      & 20.55     & 18.65     & 6.59     & 22.72      \\ \cline{2-8} 
					& \small{Aug-Clean}    & 34.28     & 20.72    & 43.15     & 31.27    & 15.66    & 38.02       \\ \cline{2-8} 
					& \small{Aug-Adv(max=2)}     & 17.32     & 9.13     & 23.69       & 22.65    & 9.26    & 27.71     \\ \cline{2-8}
					& \small{Aug-Adv(max=3)}     & 16.93     & 8.73    & 22.94     & 22.21    & 8.61       & 26.65      \\ \cline{2-8} 
					& \small{Masked-Clean} &   36.55       & 21.03         & 45.11       & 31.37     & 15.13      &37.54      \\ \cline{2-8} 
					& \small{Masked-Adv(max=2)}      & 19.56     & 10.16      & 26.43     &23.48    &9.96       & 29.63      \\ \cline{2-8} 
					& \small{Masked-Adv(max=3)}   & 18.94         &9.64        & 25.42       & 24.44     & 10.70       & 30.73       \\ \hline
					\multirow{6}{*}{\small{CSCG}}        & \small{Normal-Clean}    & 42.39   & 25.77     & 53.61    &30.82   & 17.67      & 48.14      \\ \cline{2-8} 
					& \small{Normal-Adv(max=2)}      & 8.85    & 5.10     & 23.50    & 11.90    & 6.23  & 32.94    \\ \cline{2-8}
					&\small{Normal-Adv(max=3)}     & 8.95    &  5.02  & 23.44  & 12.08     & 6.15      & 32.83     \\ \cline{2-8} 
					& \small{Aug-Clean}    &35.37     & 20.22     & 49.65      &   27.99   &  15.72     &  46.01      \\ \cline{2-8} 
					& \small{Aug-Adv(max=2)}      & 15.93    & 8.44    & 31.51    &    19.86    &  10.02       &  38.25       \\ \cline{2-8} 
					& \small{Aug-Adv(max=3)}      & 15.52     & 8.01     & 30.83      &    19.55    &   9.75     &  37.87       \\ \cline{2-8} 
					& \small{Masked-Clean} & 35.39     & 20.82    & 51.42    & 29.18     & 16.39      & 46.53       \\ \cline{2-8} 
					& \small{Masked-Adv(max=2)} & 18.37   & 10.38     & 34.44    & 16.61     & 8.04     & 35.04     \\ \cline{2-8} 
					& \small{Masked-Adv(max=3)}   & 17.75   & 9.96 & 33.77      & 16.23   & 7.67    & 34.52     \\ \hline
					\multirow{6}{*}{\small{Rencos}}        & \small{Normal-Clean}  &44.0   & 25.73 &  54.02   &   33.34  & 18.65    & 43.37   \\ \cline{2-8} 
					& \small{Normal-Adv(max=2)} 		     & 40.68 &  23.09   & 29.17    & 31.02   & 15.78  & 38.84   \\ \cline{2-8}
					&\small{ Normal-Adv(max=3)}     &  40.23  &  22.69 & 48.46  &  30.55    &  15.19    &  37.90   \\ \cline{2-8} 
					& \small{Aug-Clean}    &  41.47  &  24.18   &   51.59  &  15.58   &   3.83  &   17.53     \\ \cline{2-8} 
					& \small{Aug-Adv(max=2)}      &  40.33    & 23.02     &  49.10   &  15.84   &  3.93 &   17.57    \\ \cline{2-8} 
					& \small{Aug-Adv(max=3)}      &  40.27    &  22.93    &   48.84   & 15.93      & 3.98       &  17.68     \\ \cline{2-8} 
					& \small{Masked-Clean} &   43.73  & 25.26 &  52.53 &   33.05  &    18.25   &  42.58      \\ \cline{2-8} 
					& \small{Masked-Adv(max=2)} & 43.51  & 24.88  & 51.69     & 32.46    & 17.39  &  41.09    \\ \cline{2-8} 
					& \small{Masked-Adv(max=3)}   & 43.48  & 24.86    &  51.62    &  32.32  &  17.21  &   40.77   \\ \hline
			\end{tabular}
		\end{adjustbox}
			\label{Table:11}
		\end{center}
	}
\end{table}

\subsection{Human evaluation}
\secondrev{To complement the above objective metrics, we also conduct a human evaluation to further assess the quality of the comments generated by the masked training method, data augmentation and normal training method}. Generally, we follow the evaluation settings from the previous work~\cite{2016Summarizing, wei2020retrieve}. Particularly, the comments are examined from three aspects, i.e., similarity, naturalness, and informativeness~\cite{wei2020retrieve}. Similarity refers to how similar the generated comment is to the reference comment; naturalness measures the grammaticality and fluency; informativeness focuses on the content delivery from code snippet to the generated comments. For each of the five base models, we randomly select 20 Java code snippets and 20 Python code snippets respectively, and use the two adversarial training methods to generate comments. \secondrev{We obtain 600 generated comments and 200 references in total. To facilitate comparison, for each code snippet we construct a tuple consisting of a reference and three generated comments; we obtain 200 tuples accordingly.}

We ask six graduate students studying in the Software Engineering programme to participate in the evaluation, all of whom have at least three years of programming experience in both Java and Python, and are professionally proficient in English. The subjects are evenly divided into two groups each of which has 3 students. The 200 comment triples are also evenly divided to two parts and assigned to the two groups randomly, with each group of 100 triples. Participants manually inspect the 100 generated comment triples as well as the code snippets, and rate them independently, which means that each comment triple is examined by three individuals. The grades are given in the Likert scale ranging from 1 to 5, corresponding to `very poor', `poor', `neutral', `good', `very good' respectively where a higher value indicates a better quality. To be fair, the labels of the generator information in the triples are removed. Table~\ref{Table:12} and Table~\ref{Table:13} show the statistics of the collected results. We can observe that, on both datasets and for all the three aspects, \secondrev{the average scores of the comments generated by the masked training methods are consistently higher than those generated by the data augmentation method and normal training method}. Moreover, a majority of comments generated by the masked training method, receive scores above 3.

\begin{table}[h]
	\color{black}{
		\centering
	\caption{The evaluation results of the generated comments}
	\begin{adjustbox}{width=\columnwidth,center}
	\begin{tabular}{|c|c|c|c|c|c|c|c|}
		\hline
		\multirow{2}{*}{}                                                            & \multirow{2}{*}{\textbf{Score}} & \multicolumn{3}{c|}{\textbf{Java}}                                & \multicolumn{3}{c|}{\textbf{Python}}                              \\ \cline{3-8} 
		&                                  & \textbf{Similarity} & \textbf{Naturalness} & \textbf{Informativeness} & \textbf{Similarity} & \textbf{Naturalness} & \textbf{Informativeness} \\ \hline

		\multirow{5}{*}{\begin{tabular}[c]{@{}c@{}}Normal\\ training\end{tabular}} 
		& 5              &11(3.67\%)                & 35(11.67\%)                  & 12(4\%)                   & 25(8.33\%)                  & 63(21\%)                  & 20(6.67\%)                   \\ 
		& 4                                & 34(11.33\%)                 & 95(31.67\%)                & 40(13.33\%)                  & 58(19.33\%)                  & 103(34.33\%)                   & 45(15\%)                   \\
		
		& 3                                & 67(22.33\%)                  & 83(27.67\%)                   & 58(19.33\%)                   & 87(29\%)                  & 95(31.67\%)                   & 64(21.33\%)                   \\
		& 2                                & 115(38.33\%)                 & 59(19.67\%)                  & 52(17.33\%)                   & 68(22.67\%)                  & 41(13.67\%)                   & 41(13.67\%)                   \\
		& 1                                & 73(24.33\%)                  & 28(9.33\%)                   & 138(46\%)                  & 62(20.67\%)                 & 8(2.67\%)                   & 130(43.33 \%)                  \\ \hline

		\multirow{5}{*}{\begin{tabular}[c]{@{}c@{}}Data\\ augmentation\end{tabular}} 	& 5                                & 34(11.33\%)                  & 178(59.33\%)                  & 33(11\%)                   & 43(14.33\%)                  & 204(68\%)                  & 48(16\%)                   \\ 
			& 4                                & 89(29.67\%)                 & 56(18.67\%)                & 41(13.67\%)                  & 58(19.33\%)                  & 34(11.33\%)                   & 32(10.67\%)                   \\
		
			& 3                                & 52(17.33\%)                  & 33(11\%)                   & 67(22.33\%)                   & 28(9.33\%)                  & 12(4\%)                   & 17(5.67\%)                   \\
			& 2                                & 58(19.33\%)                 & 22(7.33\%)                  & 55(18.33\%)                   & 70(23.33\%)                  & 20(6.67\%)                   & 40(13.33\%)                   \\
		 & 1                                & 67(22.33\%)                  & 11(3.67\%)                   & 104(34.67\%)                  & 101(33.67\%)                 & 30(10\%)                   & 163(54.33\%)                  \\ \hline
		\multirow{5}{*}{\begin{tabular}[c]{@{}c@{}}Masked\\ training\end{tabular}}   
		& 5                                & 42(14\%)                  & 213(71\%)                  & 42(14\%)                   & 51(17\%)                  & 210(70\%)                  & 57(19\%)                   \\ 
		& 4                                & 191(63.67\%)                 & 43(14.33\%)                   & 112(37.33\%)                  & 167(55.67\%)                 & 52(17.33\%)                   & 118(39.33\%)                  \\
		
		& 3                                & 48(16\%)                  & 32(10.67\%)                   & 83(27.67\%)                   & 52(17.33\%)                  & 15(5\%)                   & 74(24.67\%)                   \\
		& 2                                & 14(4.67\%)                  & 10(3.33\%)                    & 35(11.67\%)                  & 20(6.67\%)                  & 13(4.33\%)                   & 34(11.33\%)                   \\
		 & 1                                & 5(1.67\%)                   & 2(0.67\%)                    & 28(9.33\%)                   & 10(3.33\%)                   & 10(3.33\%)                  & 17(5.67\%)                   \\	\hline
	\end{tabular}
\end{adjustbox}
\label{Table:12}}
\end{table}

\begin{table}[h]
	\color{black}{
	\centering
	\caption{The average results of the generated comments}
	\begin{adjustbox}{width=\columnwidth,center}

	\begin{tabular}{|c|c|c|c|c|c|c|}
		\hline
		\multirow{2}{*}{}                                           & \multicolumn{3}{c|}{\textbf{Java}}                                & \multicolumn{3}{c|}{\textbf{Python}}                              \\ \cline{2-7} 
		& \textbf{Similarity} & \textbf{Naturalness} & \textbf{Informative} & \textbf{Similarity} & \textbf{Naturalness} & \textbf{Informative} \\ \hline
		\begin{tabular}[c]{@{}c@{}}Normal\\ training\end{tabular}   & 2.32                & 3.17                 & 2.12                 & 2.72                & 3.67                 & 2.28                 \\ \hline
		\begin{tabular}[c]{@{}c@{}}Data\\ augmentation\end{tabular} & 2.88                & 4.23                 & 2.48                 & 2.57                & 4.21                 & 2.21                 \\ \hline
		\begin{tabular}[c]{@{}c@{}}Masked\\ training\end{tabular}   & 3.84                & 4.52                 & 3.35                 & 3.76                & 4.46                 & 3.55                 \\ \hline
	\end{tabular}

	\end{adjustbox}
	\label{Table:13}}
\end{table}
\subsection{Examples}
\begin{figure}[!t]
	\centering
	\includegraphics[scale=0.6]{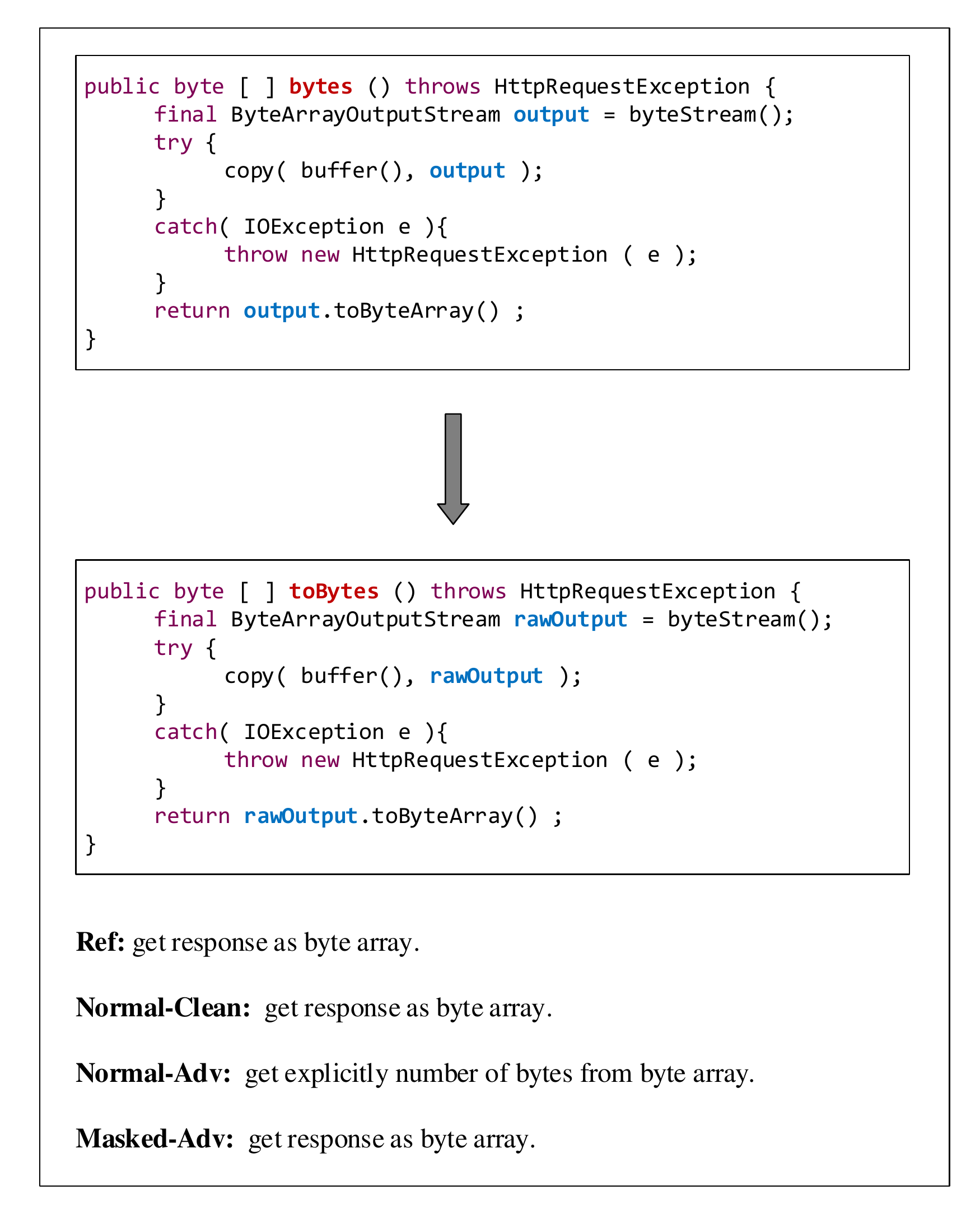}
	\caption{Examples and corresponding adversarial examples  generated by \toolname,  where `Ref' is the reference comment, `Normal-Clean' is the result of the clean example on the standard training model, `Nor-Adv' is the result of the adversarial example  on the standard training model and `Masked-Adv' is the result of the adversarial example on the masked training model.}
	\label{fig:example1}
\end{figure}

\begin{figure}[!t]
	\centering
	\includegraphics[scale=0.6]{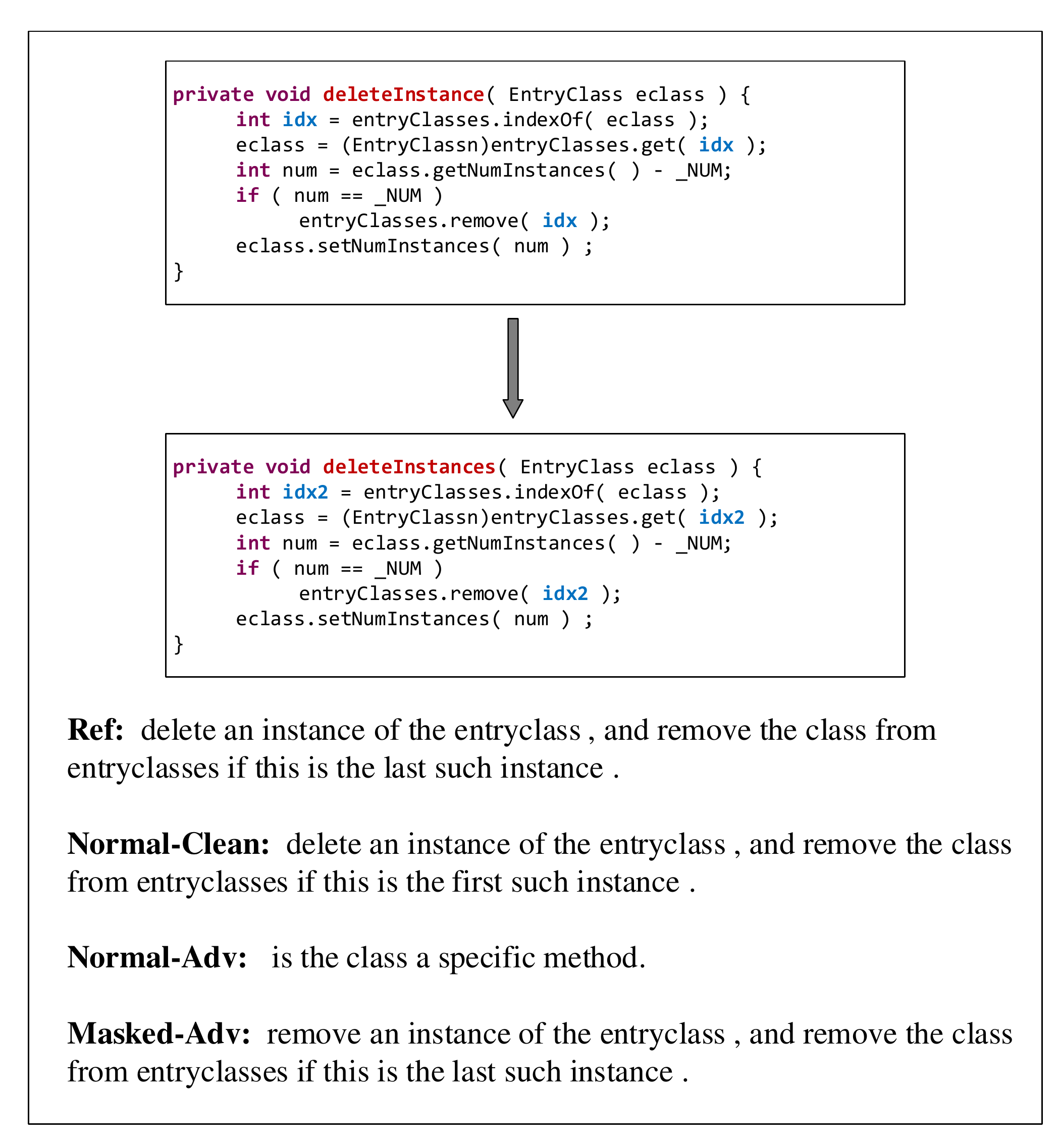}
	\caption{An example and corresponding adversarial example  generated by \toolname,  where `Ref' is the reference comment, `Normal-Clean' is the result of the clean example on the standard training model, `Nor-Adv' is the result of the adversarial example  on the standard training model and `Masked-Adv' is the result of the adversarial example on the masked training model.}
	\label{fig:example2}
\end{figure}

\begin{figure}[!t]
	\centering
	\includegraphics[scale=0.6]{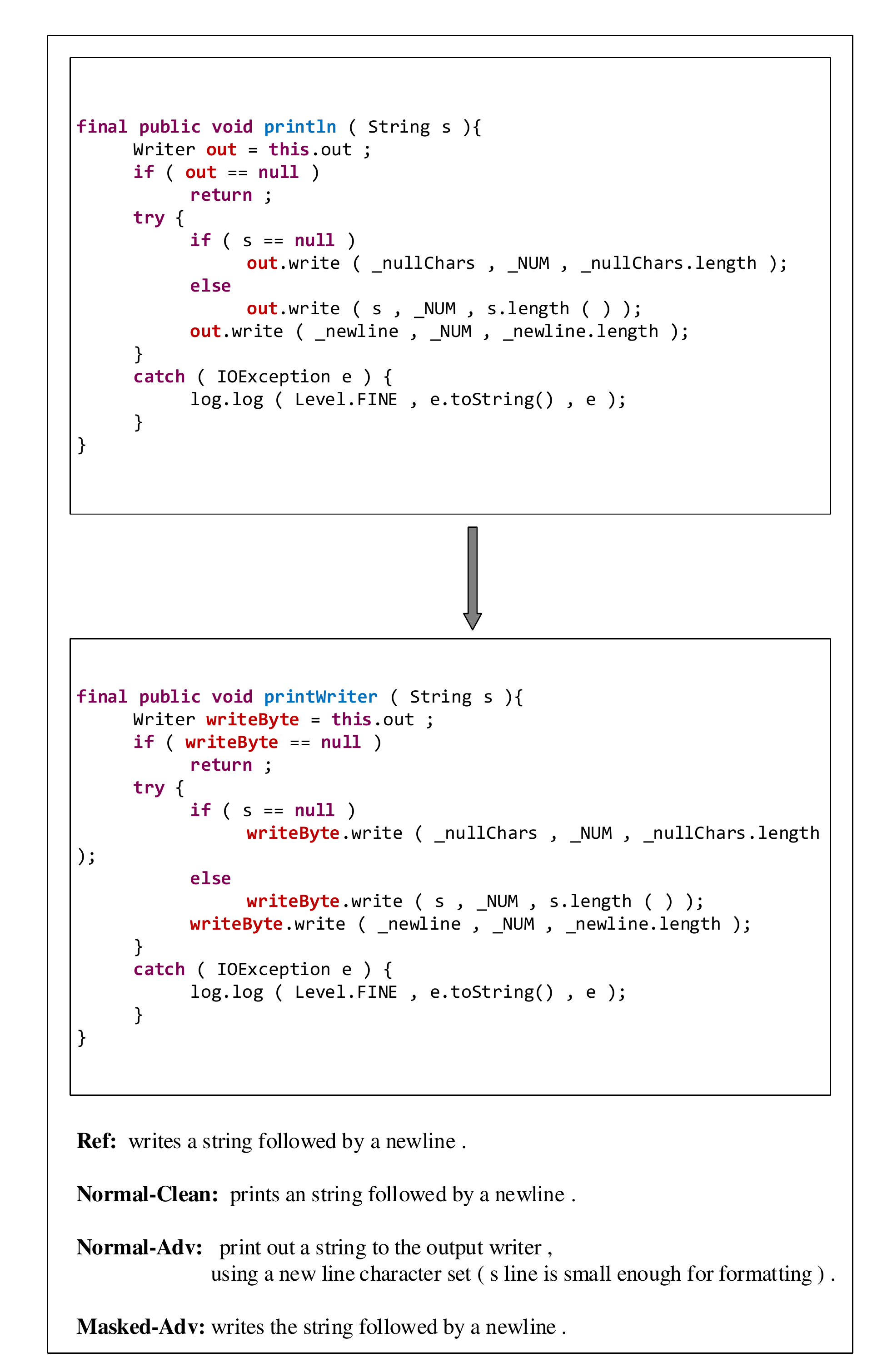}
	\caption{An example and corresponding adversarial example  generated by \toolname,  where `Ref' is the reference comment, `Normal-Clean' is the result of the clean example on the standard training model, `Nor-Adv' is the result of the adversarial example  on the standard training model and `Masked-Adv' is the result of the adversarial example on the masked training model.}
	\label{fig:example3}
\end{figure}

\begin{figure}[!t]
	\centering
	\includegraphics[scale=0.6]{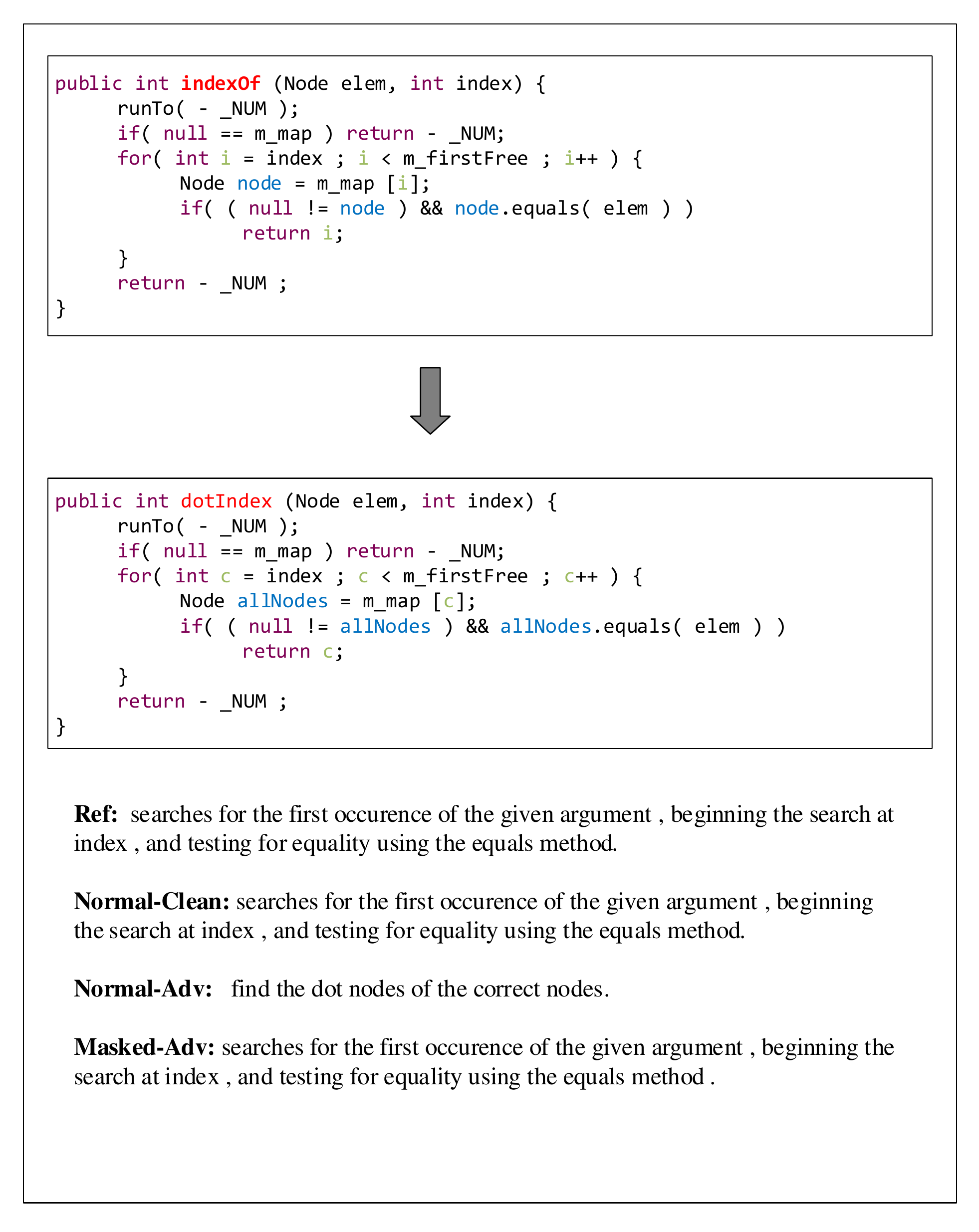}
	\caption{Examples and corresponding adversarial examples  generated by \toolname,  where `Ref' is the reference comment, `Normal-Clean' is the result of the clean example on the standard training model, `Nor-Adv' is the result of the adversarial example  on the standard training model and `Masked-Adv' is the result of the adversarial example on the masked training model.}
	\label{fig:example4}
\end{figure}

For qualitative analysis, Figure~\ref{fig:example1}--Figure~\ref{fig:example4} show some examples where `Ref' refers to the reference comment, `Normal-Clean' refers to the result of the clean example on the standard training model, `Nor-Adv' refers to the result of the adversarial example  on the standard training model, and `Masked-Adv' refers to the result of the adversarial example on the masked training model. 

We can see 
	that, the adversarial examples generated by {\toolname} are very similar to the original code snippet, indicating that our approach can generate high-quality adversarial examples preserving  original syntax, semantics and functionality. We also find that, although the standard training model (`Normal-Clean') performed well on original examples, the quality of the generated comments on the adversarial examples are poor (`Normal-AdV').
On the other hand, the masked training method can effectively defense against attacks and generate
the closest comment (`Masked-Adv') to the reference (`Ref').
%

\section{Threats to Validity}\label{Threats to Validity}
Threats to internal validity are related to internal factors that could have influenced the results. One threat that may make the results statistically unstable is the randomness from Step 2 in the adversarial attack method. In this step, we randomly select $K$ candidates for each identifier. To mitigate this, we randomly sample the $K$ candidates several times and have confirmed that our method outperforms the baselines consistently. Another threat is related to the errors introduced in the implementation. To minimize these, we have double-checked and peer-reviewed our code and repeatedly conducted the baseline methods to ensure the fairness of the results.

External validity concerns the generalizability of the results on the datasets other than the ones used in the experiments \cite{feldt10}. Indeed, in our approach, we only focus on whether code comment generation models are vulnerable to adversarial examples and how to improve the robustness of different models for Java and Python methods. However, our approach is essentially independent of specific programming languages. Note that in the adversarial attack method we intend to find the most important tokens with respect to the model, and in the masked training, we only mask the programmer-defined  identifiers in the method. Both of them can be easily applied to other datasets. Another threat originates from replacing the programmer-defined  identifiers with meaningless labels as done in CODENN \cite{2016Summarizing}, which could invalidate \toolname. However, in general, most of the work includes  the programmer-defined identifiers as part of the input to the deep code comment generation model.

\section{Related Work} \label{sect:rel}
\noindent \textbf{Code Comment Generation.} 
Code comment generation is an essential part of the software development cycle and has attracted significant attention. Neural network based approaches have been applied to this task, which is the main focus of the discussion. Alllamanis et al. \cite{allamanis2015bimodal} adopted convolutional attention neural network to generate short and name-like comments. More recent work casts it as a seq2seq generation task and employs the encoder-decoder model as the basic architecture. Based on different code representations, these approaches can be divided into two categories, i.e., token sequence based and tree structural based approaches.

For token sequence based approaches, Iyer et al.~\cite{2016Summarizing} presented an end-to-end neural attention model using LSTMs to generate comments for C\# and SQL language. Hu et al.~\cite{Hu2018Summarizing} adopted a transfer learning method utilizing API information to comment generation.  Wei et al.~\cite{NIPS2019_8883} utilized dual learning to train a code comment generation model and code generation model simultaneously. Ahmad et al.~\cite{2020A} adopted Transformer with absolute position encoding to comment generation. 

For tree structural  based approaches, the general methodology is to encode the code as (variants of) ASTs  which are input to purpose-designed neural networks. Hu et al. \cite{hu2018deep} proposed a structure based traversal method to flatten the AST. LeClair et al.~\cite{subroutines}  aimed to combine words from code with code structure from AST. Furthermore, techniques have been put forward to  enhance the performance, e.g., approaches based on reinforcement learning \cite{2018Improving} or aided with contextual information  \cite{Yu2019Augmenting}.

\smallskip
\noindent \textbf{Adversarial Examples Generation.} 
Adversarial examples were first proposed by Szegedy et al. in image classification \cite{szegedy2013intriguing}. Their experiment shows that an imperceptible perturbation of the benign input image could cause misclassification. A plethora of generation methods have been studied for image classification, a thorough survey of which is clearly out of the scope of the current paper. Here we only mention some representational work such as 
FGSM \cite{Goodfellow2015Explaining}, 
Deepfool \cite{Moosavi2016DeepFool}, 
BIM \cite{kurakin2016adversarial}, 
JSMA \cite{papernot2016the}, and 
the C\&W method \cite{Carlini2016Towards}.

Our work is more related to adversarial example generation in the NLP area which turns out to be more challenging although the underlying principles are somewhat similar. 
Natural language texts are discrete and are more difficult to be perturbed in a meaningful way. Papernot et al.~\cite{Papernot2016Crafting} first studied the problems of adversarial examples in text by adopting FGSM. Semanta et al.~\cite{2017Towards} combined FGSM and importance of word to select the top-k words with highest importance to attack the text classification model. Similarly, Ren et al.~\cite{2019Generating} proposed PWWS which based on word saliency to attack the text classification model. Jia et al.~\cite{jia2017adversarial} added sentences to the ends of paragraphs using crowdsourcing to fool reading comprehension system. Belinkov et al.~\cite{belinkov2018synthetic} devised adversarial examples depending on natural and synthetic language errors which can fool Neural Machine Translation (NMT) system.

Comparing to the large body of work on adversarial examples for image and NLP, the corresponding work for source code
processing is in its infancy; this is especially the case for comment generation.

Bielik et al.~\cite{Pavol2020} improved  the adversarial robustness of models for the task of type inference by learning to abstain if uncertain. Zhang et al.~\cite{zhang2020generating} studied the problem for the code classification tasks where they proposed a sampling based method to generate adversarial examples. Note that this work is still of classification nature whereas our work focuses on comment generation, which is of language \emph{translation} or \emph{generation} nature.  
Yefet et al.~\cite{Yefet2019} generated adversairal example based on gradient for CODE2VEC. Ramakrishnan et al.~\cite{ramakrishnan20} and Ravichandar et al.~\cite{STRATA} both performed adversarial attacks and adversarial training on CODE2SEQ. They all concentrate on method name prediction instead of generating long comment that help programmers understand.

\smallskip
\noindent\textbf{Adversarial Defense.}
There have some relatively effective methods against adversarial attacks in NLP, which can roughly be classified as detection and model enhancement methods.
Li et al.~\cite{2018TextBugger} proposed to use a context-aware spelling check service to detect spell errors in adversarial examples. 
Pruthi et al.~\cite{pruthi2019combating}  proposed a method to combat adversarial spelling mistakes by placing a word recognition model in front of the downstream DNNs. Wang et al.~\cite{2020Defense} proposed an adversarial defense method SEM, which inserts an encoder network before the original model and trains it to eliminate  adversarial perturbations.


In addition to the detection-based defense, adversarial training as a typical model enhancement method, is also widely adopted. Javid et al.~\cite{2018HotFlip} used adversarial training to improve the robustness of text classification model. Wang et al.~\cite{wang2018robust} augmented original training dataset with adversarial examples generated by AddSentDiverse to enhance the robustness of reading comprehension models. In order to improve the robustness of text classification, Ren et al.~\cite{2019Generating} randomly selected clean examples from the training set to generate adversarial examples using PWWS and mixed them with the training dataset to conduct adversarial training. Moreover, other work such as \cite{sato2018interpretable,zhang2019Grating,zang2020word,zhang2020generating} adopted adversarial training to improve the robustness of DNN models. 

\section{Conclusion} \label{sect:conc}

In this paper, we have presented a novel approach {\toolname} to address the adversarial robustness problem of DNN models for code comment generation tasks, and demonstrated that the current mainstream code comment generation architectures are of poor robustness. Simply replacing identifiers which results in functionality-persevering and syntactically correct code snippets can degrade the performance of these representative models greatly. Experiment results show that our method can generate more effective adversarial examples on two public datasets across five mainstream code comment generation architectures.  In addition, we demonstrated that the adversarial examples generated by our method had better transferability. To improve robustness, we have also proposed a novel training method.
Our experimental results showed that this training method can achieve better performance in the code comment generation setting compared to the data augmentation method  which has widely been used to improve robustness.

In the future, we plan to extend the existing framework and include more sophisticated, structure-rewriting based adversarial example generation techniques. More generally, we plan to explore the robustness issues of machine learning models for  
other software engineering tasks.

\begin{acks}
This work was partially supported by the National Natural Science Foundation of China (NSFC, No.\ 61972197), the Natural Science Foundation of Jiangsu Province (No.\ BK20201292), the Collaborative Innovation Center of Novel Software Technology and Industrialization, and the Qing Lan Project. T.\ Chen is partially supported by Birkbeck BEI School Project (ARTEFACT), NSFC grant (No.\ 61872340 and No.\ 62072309), UK EPSRC grant (EP/P00430X/1), Guangdong Science and Technology Department grant (No.\ 2018B010107004) and an oversea grant from the State Key
Laboratory of Novel Software Technology, Nanjing University (KFKT2018A16).
\end{acks}

\bibliographystyle{Reference}
\bibliography{acmart}
 
%
\end{document}